\documentclass[journal=jacsat,manuscript=article]{achemso}
\usepackage[T1]{fontenc}
\usepackage{graphicx}
\usepackage{dcolumn}
\usepackage{bm}
\usepackage{hyperref} 
\usepackage{xcolor}
\usepackage{siunitx}
\usepackage{xspace}
\usepackage{chemformula}
\usepackage{nameref}
\usepackage{booktabs}
\usepackage{caption}
\usepackage{subcaption}
\usepackage{comment}
\hypersetup{pdfborder=0 0 0,colorlinks=true,citecolor=black,linkcolor=black}
\usepackage{amsmath,amssymb}
\usepackage{nameref}
\usepackage{tabularx}
\usepackage{siunitx} 
\mciteErrorOnUnknownfalse
\input epsf

\newcommand\THEOSMARVEL{Theory and Simulation of Materials (THEOS), and National Centre for Computational Design and Discovery of Novel Materials (MARVEL), {\'E}cole Polytechnique F{\'e}d{\'e}rale de Lausanne, 1015 Lausanne, Switzerland}
\newcommand\PSI{Laboratory for Materials Simulations, Paul Scherrer Institut (PSI), 5232 Villigen, Switzerland}
\newcommand\JENA{Institut für Festkörpertheorie und -optik, Friedrich-Schiller-Universität Jena, Jena, Germany}
\newcommand\TORVERGATA{ Dipartimento di Fisica, Universit{\`a} di Roma Tor Vergata, Roma, Italy}
\newcommand\ROMATRE{Dipartimento di Ingegneria Industriale, Elettronica, e Meccanica,  Universit{\`a} Roma Tre, Roma, Italy}
\newcommand\BICOCCA{Dipartimento di Scienza dei Materiali, Universit{\`a} degli studi di Milano Bicocca, Piazza dell'Ateneo Nuovo 1, 20126 Milano, Italy}
\newcommand\BIQUTE{Bicocca Quantum Technologies (BiQuTe) Centre, I-20126 Milano, Italy}

\title{Quasiparticle effects and strong excitonic features in exfoliable 1D semiconducting materials}

\author{Simone Grillo}
\affiliation{\TORVERGATA}
\email{simone.grillo@roma2.infn.it}
\author{Chiara Cignarella}
\affiliation{\THEOSMARVEL}
\email{ccignare@uni-bremen.de}
\affiliation{\THEOSMARVEL}
\author{Friedhelm Bechstedt}
\affiliation{\JENA}
\author{Paola Gori}
\affiliation{\ROMATRE}
\author{Maurizia Palummo}
\affiliation{\TORVERGATA}
\author{Davide Campi}
\affiliation{\BICOCCA}
\alsoaffiliation{\BIQUTE}
\author{Nicola Marzari}
\affiliation{\THEOSMARVEL}
\alsoaffiliation{\PSI}
\author{Olivia Pulci}
\affiliation{\TORVERGATA}
\email{olivia.pulci@roma2.infn.it}

\date{\today}

\begin{document}
\maketitle

\begin{abstract}
We report a comprehensive first-principles study of the electronic and optical properties of recently identified exfoliable one-dimensional semiconducting materials, focusing on chalcogenide-based atomic chains derived from van der Waals-bonded bulk crystals. Specifically, we investigate covalently bonded S$_3$ and Te$_3$ chains, and polar-bonded As$_2$S$_3$ and Bi$_2$Te$_3$ chains, using a fully first-principles approach that combines density-functional theory (DFT), density-functional perturbation theory (DFPT), and many-body perturbation theory within the $GW$ approximation and Bethe–Salpeter equation (BSE).
Our vibrational analysis shows that freestanding isolated wires remain dynamically stable, with the zone-center optical phonon modes leading to infrared activity.
The main finding of this study is the presence of very strong exciton binding energies (1–3 eV), which make these novel 1D materials ideal platforms for room-temperature excitonic applications. Interestingly, the exciton character remains Wannier–Mott-like, as indicated by average electron–hole separations larger than the lattice constant.
Notably, the optical gaps of these materials span a wide range --- from infrared (0.8 eV, Bi$_2$Te$_3$), through visible spectrum (yellow: 2.17 eV, Te$_3$; blue: 2.71 eV, As$_2$S$_3$), up to ultraviolet (4.07 eV, S$_3$) --- highlighting their versatility for broadband optoelectronic applications.
Our results offer a detailed, many-body perspective on the optoelectronic behavior of these low-dimensional materials and underscore their potential for applications in next-generation nanoscale optoelectronic devices.
\end{abstract}


One-dimensional (1D) materials, such as nanotubes and nanowires, represent a rapidly evolving frontier in advanced materials research, owing to their exceptional electronic, optical, and mechanical properties. Their reduced dimensionality gives rise to distinct quantum phenomena that profoundly influence charge transport, light–matter interactions, and coupling with the surrounding environment. These effects position 1D materials as highly promising candidates for a wide range of technological applications, including nanoelectronics, spintronics, and sensing technologies \cite{meng2022one, balandin2022one, zhu2023machine}.
Among the most exciting directions is their use in optoelectronics, where the strong optical response of 1D systems --- characterized by enhanced light absorption and tightly bound excitons --- opens new avenues for the design of efficient photodetectors, lasers, and next-generation solar cells.
A particularly promising strategy for accessing single nanowires has emerged in recent years, based on the exfoliation of strongly anisotropic three-dimensional (3D) crystals composed of covalently bonded inorganic chains held together by weak inter-chain interactions \cite{balandin2022one, meng2022one}. In these materials, individual 1D wires are arranged in a crystal lattice via van der Waals (vdW) forces, allowing for their isolation through mechanical or chemical exfoliation techniques \cite{stolyarov2016breakdown, geremew2018current, island2017electronics, lipatov2018quasi}, in analogy with the well-established methods developed for two-dimensional (2D) materials.

\noindent Unlike carbon nanotubes, exfoliated wires offer well-defined and reproducible electronic and optical properties, making them highly attractive for nanoelectronic applications. Their structural uniformity and reduced edge scattering have demonstrated significant promise as fundamental components in next-generation electronic devices, particularly field-effect transistors (FETs) \cite{stolyarov2016breakdown, empante2019low, geremew2018current, jeong2024tailoring}. Moreover, their anisotropic bonding and the absence of dangling bonds perpendicular to the wire axis --- exemplified in materials such as BiSI, BiSeI, Sb$_2$Se$_3$, and Bi$_2$Se$_3$ --- can be used to enhance charge carrier lifetimes and minimize nonradiative recombination, thereby boosting the performance of high-efficiency photo-conversion devices \cite{zhou2015thin, ganose2016relativistic}.

\noindent Computational high-throughput (HT) screening offers a powerful tool to discover novel one-dimensional wires.
Such studies have been effective in identifying 3D materials suitable for exfoliation in atomically thin layers \cite{lebegue2013two, ashton2017topology, choudhary2017high, cheon2017data, mounet2018two, haastrup2018computational, larsen2019definition, campi2023expansion}, and recently similar approaches have been applied to identify one-dimensional systems \cite{shang2020atomic, zhu2021spectrum, moustafa2022computational, moustafa2023hundreds,zhu2023machine}.
Recently, a large dataset of 1D systems that can be exfoliated from weakly-bonded crystals was built \cite{cignarella2024searching, campi1D}, screening three different databases (COD \cite{gravzulis2012crystallography}, ICSD \cite{karlsruhe2019inorganic, bergerhoff1983inorganic, villars1998linus}, and Pauling File \cite{pauling}) containing all structures experimentally reported in their 3D form.

\noindent In this study, we target electronic and optical properties of exfoliable 1D wires promising for future technological applications, with a particular focus on optoelectronic applications. Up to date, investigations on feasibly exfoliable 1D materials, using state-of-the-art MBPT methods are fairly sparce \cite{varsano2017carbon, andharia2018exfoliation, smolenski2025large}.
Here, we aim to predict novel electronical and optical properties and understand them in terms of dimensionality, enhanced electron-electron interaction, and quantum confinement.
We select promising materials from the above database of exfoliable 1D systems, screening for wires found mechanically stable from the calculated phonon dispersion, with maximum two atomic species per unit cell, and with semiconducting behavior with a DFT direct band gap below 3 eV. Following these guidelines, we choose four chalcogen-based materials, namely S$_3$, Te$_3$, As$_2$S$_3$, Bi$_2$Te$_3$\footnote{The nomenclature used corresponds to the number of atoms of each atomic species in the unit cell.}

\noindent To tackle these materials and the calculation of their properties, we employ density-functional theory (DFT) and many-body perturbation theory (MBPT), including $G_0W_0$, eigenvalue self-consistent $evGW$ and the Bethe-Salpeter equation (BSE). Due of the significant confinement of the inhomogenous electron gas and the strong reduction of its screening, Coulomb interactions are enhanced and, consequently, many-body exchange-correlation effects become more dramatic than in nearly-free-electron-like 3D materials. This holds especially for electronic excitations. \\

\section{Results and Discussion}
\subsection{Structural and vibrational properties}

\begin{figure*}[h!]
  \centering
  \begin{tabular}{@{}c@{}}
    \includegraphics[scale=0.35]{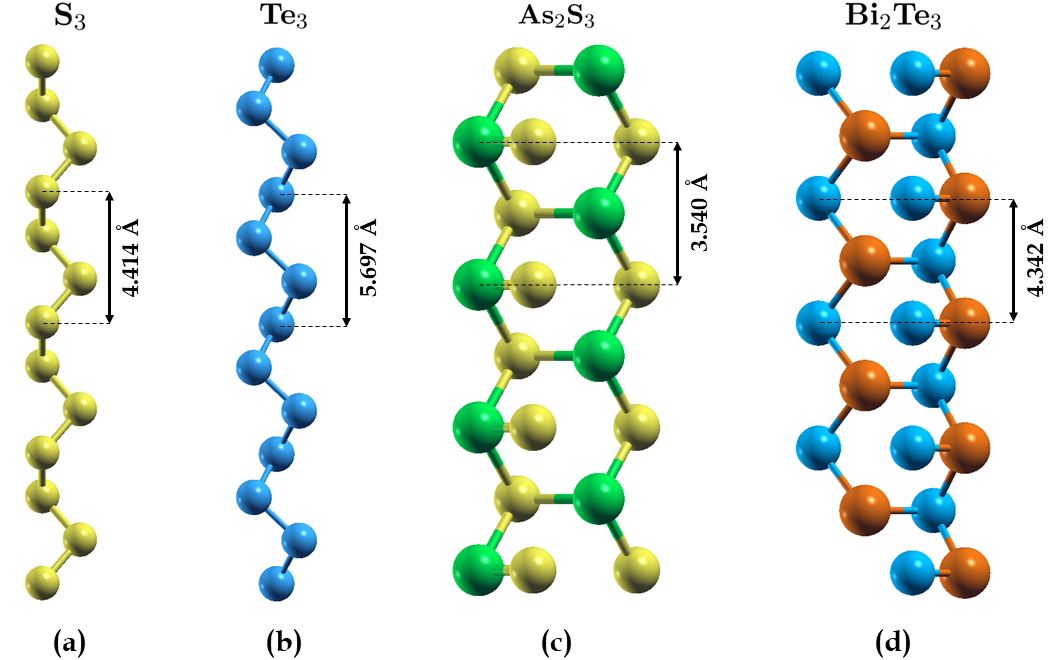} \\[\abovecaptionskip]
  \end{tabular}
  \caption{Optimized geometric structures of $(\mathbf{a})$ $\mathrm{S}_3$, $(\mathbf{b})$ $\mathrm{Te}_3$, $(\mathbf{c})$ $\mathrm{As}_2\mathrm{S}_3$ and $(\mathbf{d})$ $\mathrm{Bi}_2\mathrm{Te}_3$. The optimized lattice parameters of the unit cells are also displayed.}\label{1D_structures}
\end{figure*}

We begin our analysis by characterizing the structural properties of the 1D materials investigated in this work. These have been obtained through HT exfoliation of weakly-bonded 3D parent compounds \cite{cignarella2024searching}. The corresponding 3D systems were sourced from databases containing experimentally reported structures only. In particular, we investigate chain structures of S$_3$, Te$_3$, As$_2$S$_3$, and Bi$_2$Te$_3$ (see Fig.~\ref{1D_structures}), which can be exfoliated from \href{https://mc3d.materialscloud.org/#/details/mc3d-64810/pbe-v1}{S$_9$} \cite{pauling_file_ref_id_158251} (S$_3$), 
\href{https://mc3d.materialscloud.org/#/details/mc3d-51852/pbe-v1}{$\alpha$-Te} \cite{pauling_file_ref_id_21540, bradley1924crystal} (Te$_3$), \href{https://mc3d.materialscloud.org/#/details/mc3d-62793/pbe-v1}{As$_8$S$_{12}$} \cite{pauling_file_ref_id_98484} (As$_2$S$_3$) and \href{https://mc3d.materialscloud.org/#/details/mc3d-78106/pbesol-v1}{Bi$_4$Te$_6$} \cite{pauling_file_ref_id_219560} (Bi$_2$Te$_3$).
We perform on each material an additional full-structure relaxation to obtain the optimized lattice parameters and atomic positions using the chosen XC functional. The resulting values are reported in Table~\ref{tab1_1D}. The structures of S$_3$ (Fig.~\ref{1D_structures}a) and Te$_3$ (Fig.~\ref{1D_structures}b) both consist of a chiral helical chain composed of three atoms in the unit cell, arranged so that the projection of the chain onto a plane perpendicular to the chain axis forms an equilateral triangle. 
Such a structure is typical of chalcogen elements, even for higher dimensional arrangements. The unit cell of As$_2$S$_3$ (Fig.~\ref{1D_structures}c) consists of two As atoms and three S atoms, which are alternately bonded to form a network of non-planar hexagons. 
Additionally, there is an extra S atom attached to each hexagon (through an As atom), protruding towards its center. The structure of Bi$_2$Te$_3$ (Fig.~\ref{1D_structures}d) is specular to that of As$_2$S$_3$, with Bi and Te replacing As and S, respectively. The described geometries give rise to minima on the total energy surface, thereby indicating the statically energetic stability.

\begin{table*}[htbp!]
\normalsize
\renewcommand{\arraystretch}{1.5}
\caption{Optimized lattice parameter $(a)$ of the systems studied, obtained by full-structure relaxation using a GGA-PBE XC functional, together with the calculated lowest electronic gap $(E^{\mathrm{DFT}}_{g})$ at the DFT level. In the case of indirect band gap, the corresponding direct band gap is also reported in square brackets. SOC was
included.}
\begin{tabularx}{\textwidth}{XXXXX}
    \hline
    & \textbf{S}$_3$ & \textbf{Te}$_3$ & \textbf{As}$_2$\textbf{S}$_3$ & \textbf{Bi}$_2$\textbf{Te}$_3$ \\ \hline
    $a \,$ (\si{\angstrom}) & 4.414 & 5.697 & 3.540 & 4.342 \\
    $E^{\mathrm{DFT}}_{g} \,$ (eV) & 2.66 [2.76] & 1.43 [1.47] & 1.16 [1.26] & 0.42 [0.43] \\
    \hline
\end{tabularx}
\label{tab1_1D}
\end{table*}

\begin{figure*}[h!]
   \centering
\includegraphics[scale=0.5]{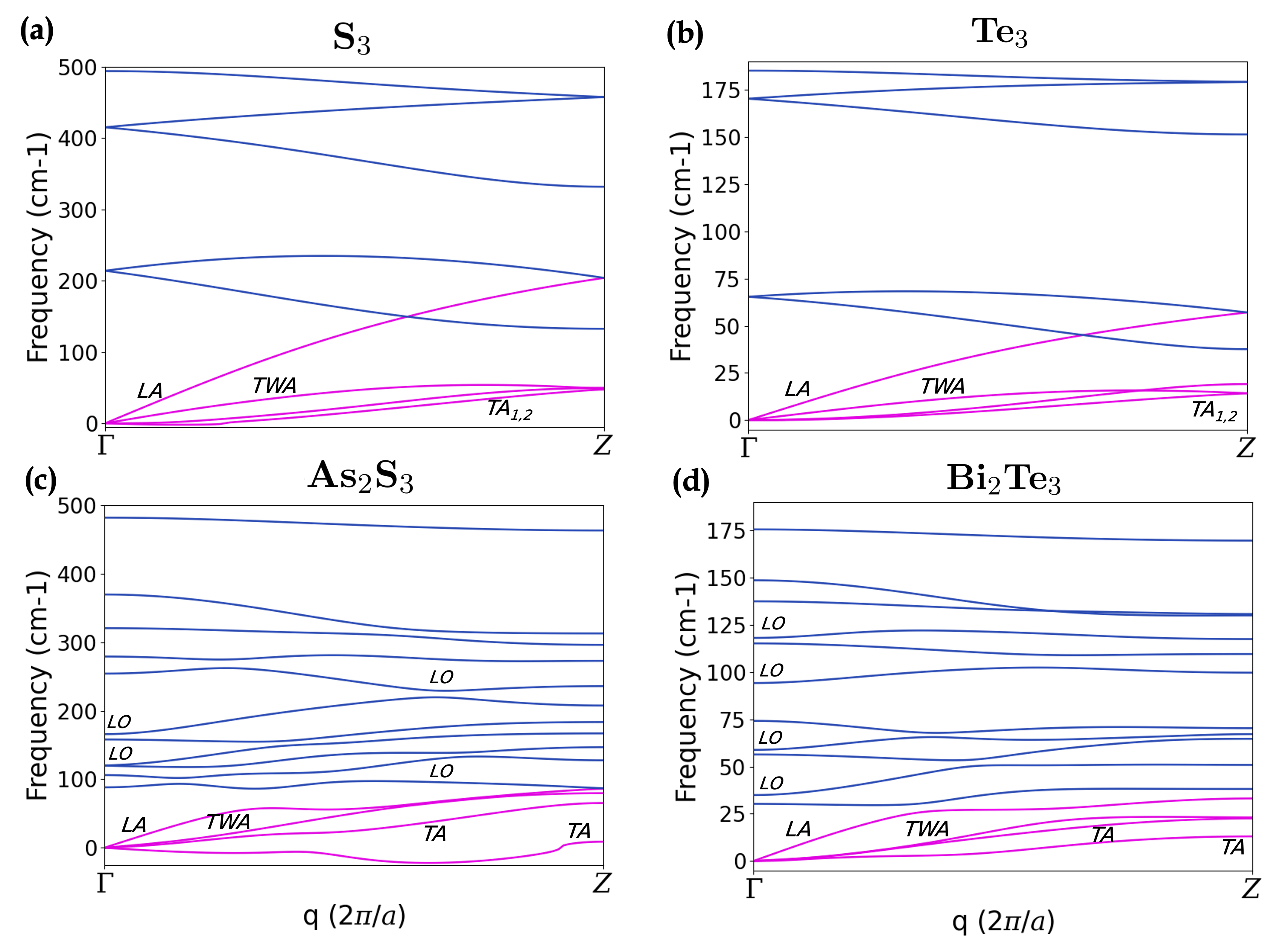} \\
\includegraphics[scale=0.35]{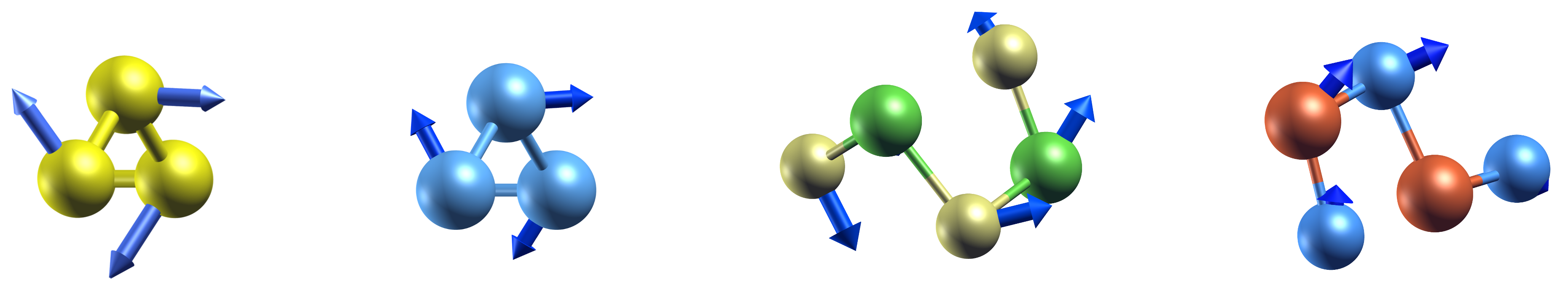}
      \caption{[Top] Phonon band structures of $(\mathbf{a})$ $\mathrm{S}_3$, $(\mathbf{b})$ $\mathrm{Te}_3$, $(\mathbf{c})$ $\mathrm{As}_2\mathrm{S}_3$ and $(\mathbf{d})$ $\mathrm{Bi}_2\mathrm{Te}_3$, calculated at harmonic level within DFPT. [Bottom] Top-view of (from left to right) $\mathrm{S}_3$, $\mathrm{Te}_3$, $\mathrm{As}_2\mathrm{S}_3$, $\mathrm{Bi}_2\mathrm{Te}_3$ with arrows indicating the displacements produced by TWA at $\Gamma$.}\label{fig:phonons}
\end{figure*}

\noindent The equilibrium geometries obtained (see Table~\ref{tab1_1D}) are used as starting structures for the investigation of the vibrational properties. We compute the phonon dispersion curves along the  1D Brillouin Zone (BZ), between the center $\Gamma$ and the boundary point $Z$, using finer \textbf{k}- and \textbf{q}-points grids, as well as larger vacuum space and stricter energy and forces thresholds with respect to the calculation performed in the source HT study (which inherently requires a compromise between accuracy and computational cost).
The results are shown in Fig.~\ref{fig:phonons}: S$_3$ (Fig.~\ref{fig:phonons}a), Te$_3$ (Fig.~\ref{fig:phonons}b) and Bi$_2$Te$_3$ (Fig.~\ref{fig:phonons}d) exhibit no imaginary phonon frequencies along the entire BZ. Therefore, these chain-like structures are also dynamically stable in their freestanding configuration as exfoliated from their 3D counterparts. We only note a small instability in As$_2$S$_3$ (Fig.~\ref{fig:phonons}c). The lowest transverse acoustic (TA) phonon branch exhibits negative frequencies around 2/3 $\Gamma Z$. This fact may be interpreted as a tendency for a reconstruction toward a tripling of the unit cell size. However, we stress that the calculations are performed in the harmonic approximation and at T $=0$ K, therefore without accounting for the effect of anharmonicity and/or temperature, which could contribute to the stability of 1D systems \cite{romanin2021dominant,bianco2019quantum,artyukhov2014mechanically,bianco2020weak}. \\
\noindent The phonon dispersions highlight four acoustic branches going to zero at $\Gamma$, outlined in magenta in Fig.~\ref{fig:phonons}. In addition to the three translational modes, 1D wires are also invariant under rotation around their axis \cite{lin2022general, cignarella2025extreme}; this means that when completing a full rotation around it, the force acting on the wire is zero, and hence the corresponding phonon frequency is also zero. We denote this mode as twisting acoustic mode (TWA); it is illustrated in Fig.~\ref{fig:phonons} for the four materials, where this vibration appears as a rotation or twisting of the nanowire. The two acoustic branches quadratic in \textbf{q}, particularly visible in Fig.~\ref{fig:phonons}a and b, are flexural modes representing transversal vibrations of atoms along the two directions perpendicular to the wire (TA) --- similar to the ZA mode in 2D monolayers --- which generate a bending of the wire. 
The highest acoustic mode is instead linear in \textbf{q}, and corresponds to longitudinal in-line vibrations (LA). 
Among the optical modes, the highest phonon branches of S$_3$ and Te$_3$ are associated with breathing modes, where atoms vibrate outward and inward from the center of the chains. Similar vibrational patterns are observed in As$_2$S$_3$ and Bi$_2$Te$_3$ in the two second-highest optical branches, while the highest isolated branch visible in Fig.~\ref{fig:phonons}c and d represents breathing vibrations involving only the extra S (Te) attached to the hexagonal body.

\begin{figure*}[h!]
   \centering
\includegraphics[scale=0.5]{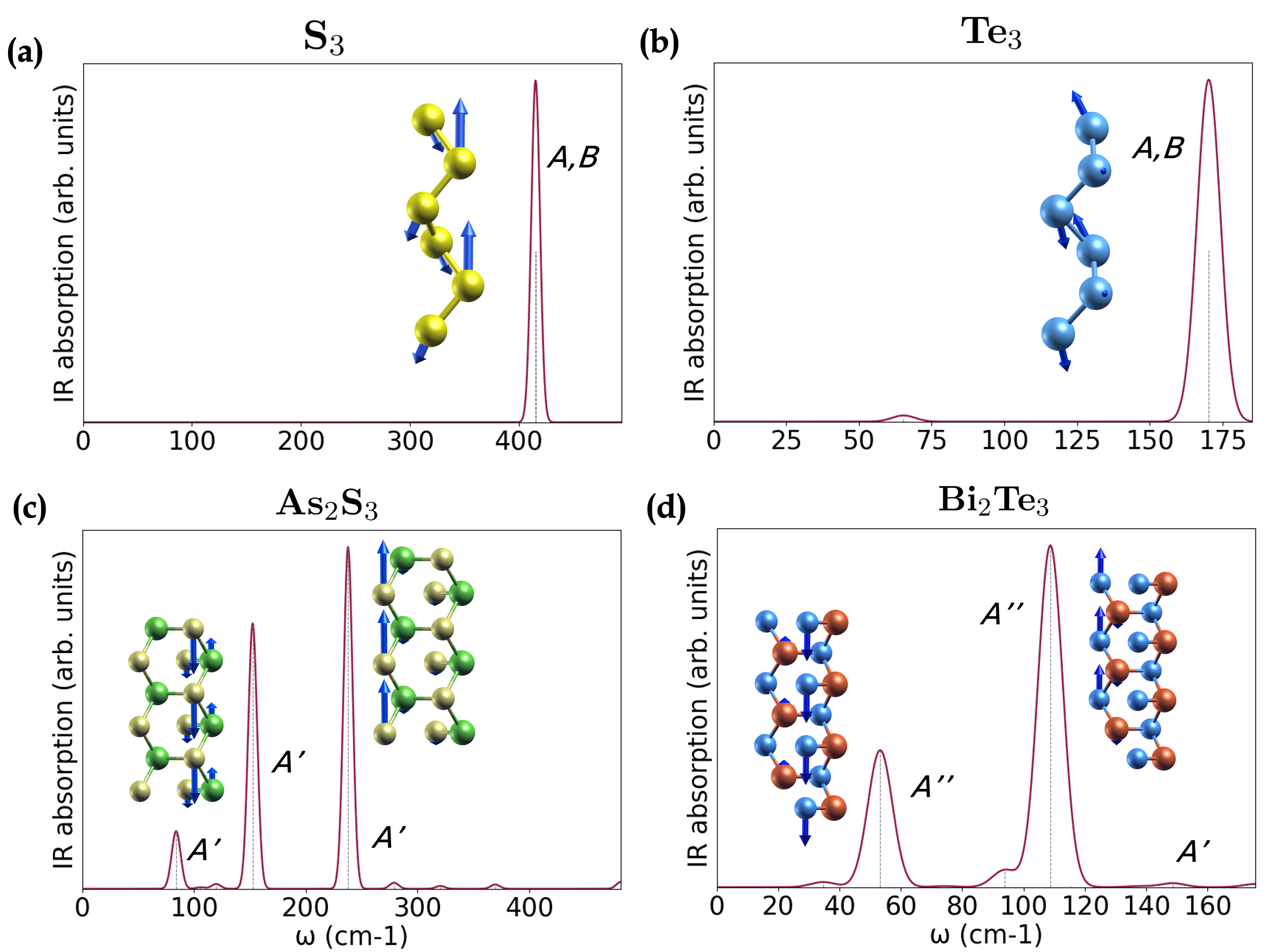}
      \caption{IR absorption spectra of $(\mathbf{a})$ $\mathrm{S}_3$, $(\mathbf{b})$ $\mathrm{Te}_3$, $(\mathbf{c})$ $\mathrm{As}_2\mathrm{S}_3$ and $(\mathbf{d})$ $\mathrm{Bi}_2\mathrm{Te}_3$, calculated at harmonic level of theory in DFPT. Insets show the corresponding vibrations of the main IR-active modes. The colors of the atoms are yellow (S), blue (Te), green (As) and red (Bi).}\label{fig:IR}
\end{figure*}

\begin{figure*}[h!]
   \centering
\includegraphics[scale=0.45]{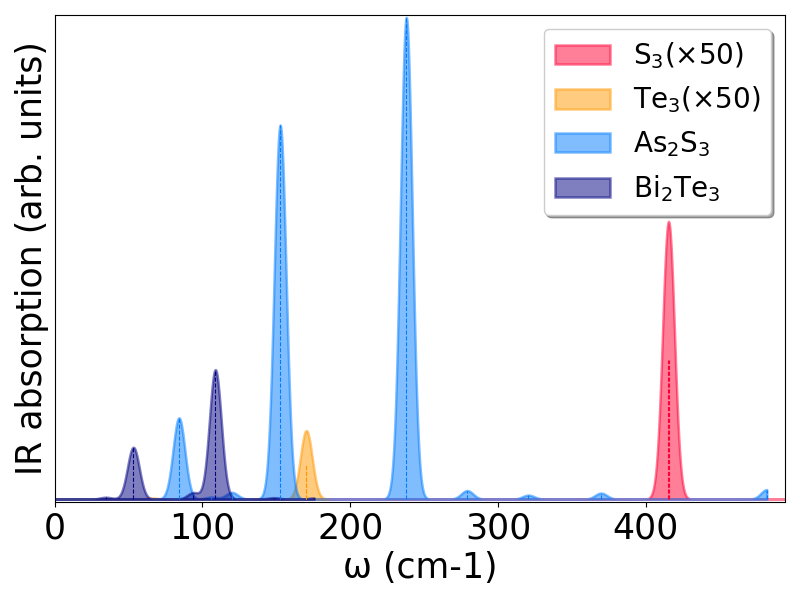}
      \caption{IR absorption peaks of the four nanowires. The IR response of S$_3$ and Te$_3$ is multiplied by a factor of 50 in the plot, as extremely weak (and otherwise indistinguishable) compared to the most prominent peaks of As$_2$S$_3$ and Bi$_2$Te$_3$. }\label{fig:IR_ensemble}
\end{figure*}

\noindent In Fig.~\ref{fig:IR} the IR absorption spectra resulting for the four wires are displayed. As$_2$S$_3$ (Fig.~\ref{fig:IR}c) and Bi$_2$Te$_2$ (Fig.~\ref{fig:IR}d) exhibit two main IR-active peaks, that couple to the atomic vibrations illustrated in the insets of the same figure. All their observable IR-active peaks correspond to LO (longitudinal optical) modes, with the oscillation strength aligned with the direction of propagation ($\hat{z}$) $\mathbf{\overline{Z}}^* \cdot \mathbf{q}_z \neq 0$ \cite{umari2001raman,bastonero2024automated}, and labeled LO in Fig.~\ref{fig:phonons}.
Such LO modes 
are expected to shift from the closest TO (transverse optical) modes visible in the phonon dispersions. This effect is due to the long-range coupling between the macroscopic electric field generated by the longitudinal vibrations and the vibration itself, which in 3D results in the so-called \textit{LO-TO splitting} between the LO and TO degenerate modes (without the field) at $\Gamma$ \cite{baroni2001phonons}. In 1D, the LO and the two TO modes are symmetrically non-equivalent and therefore not necessarily degenerate, and the long-range polar interaction produces not a splitting, but a blue-shift in frequency of the LO mode \cite{rivano2023infrared}. \\
\noindent S$_3$ (Fig.~\ref{fig:IR}a) and Te$_3$ (Fig.~\ref{fig:IR}b) chains exhibit IR spectra similar to each other, consinsting of two degenerate IR-active modes with identical vibrational patterns. In the inset, we show the \emph{A} and \emph{B} modes, respectively for S$_3$ and Te$_3$ (the vibrational patterns associated with each mode are then equivalent in both two materials).
Due to their similar geometry, only the frequencies of the phonon modes - and hence the peak positions - are (drastically) influenced by the nature of the element. For S$_3$ the peak is shifted toward higher frequencies compared to Te$_3$ by a factor 2.4, which is only slightly larger than the square root of the mass ratio $\sqrt{M_{\mathrm{Te}}/M_{\mathrm{S}}}=2$.
Interestingly, the IR response of these modes arises from vibrations along $\hat{x}$ and $\hat{y}$ directions and thus not classified as LOs: no shift or split around $\Gamma$ is expected if the long-range polar effects are included in the calculation \cite{rivano2024density,rivano2023infrared}. 
Moreover, these two materials are homopolar, \emph{i.e.}, composed of the same atomic species, hence the polarization responsible for the IR peak does not derive from a difference in electronegativity of the atomic species, but rather from their structure asymmetry. This yields to a lower charge disproportion and a weaker IR response, as shown in Fig.~\ref{fig:IR_ensemble}. Here, we compare the relative peaks of the four nanowires, normalized by their quantum volume: S$_3$ and Te$_3$ response is almost negligible as compared to that of As$_2$S$_3$ and Bi$_2$Te$_3$, the latter giving rise to the most intense peaks.
Their dipole coupling makes their intensity by two to three orders of magnitude more intense compared to the homopolar cases.

\noindent  The IR peaks in As$_2$S$_3$ and Bi$_2$Te$_3$ spectra in Fig.~\ref{fig:IR}\footnote{A slight frequency misalignment in Bi$_2$Te$_3$ between IR and phonon dispersions arises from applying different ASRs during post-processing, since the same 1D-ASRs with full Born-Huang conditions \cite{lin2022general} are not implemented in the 'dynmat' module.} appear at zone-center frequencies much below the uppermost S$_3$- or Te$_3$-dominated frequencies around 500 and $175 \,$ cm$^{-1}$, respectively (see also Fig.~\ref{fig:phonons}).

\subsection{Electronic and optical properties}

\begin{figure*}[h!]
  \centering
  \begin{tabular}{cc}
    \includegraphics[scale=0.45]{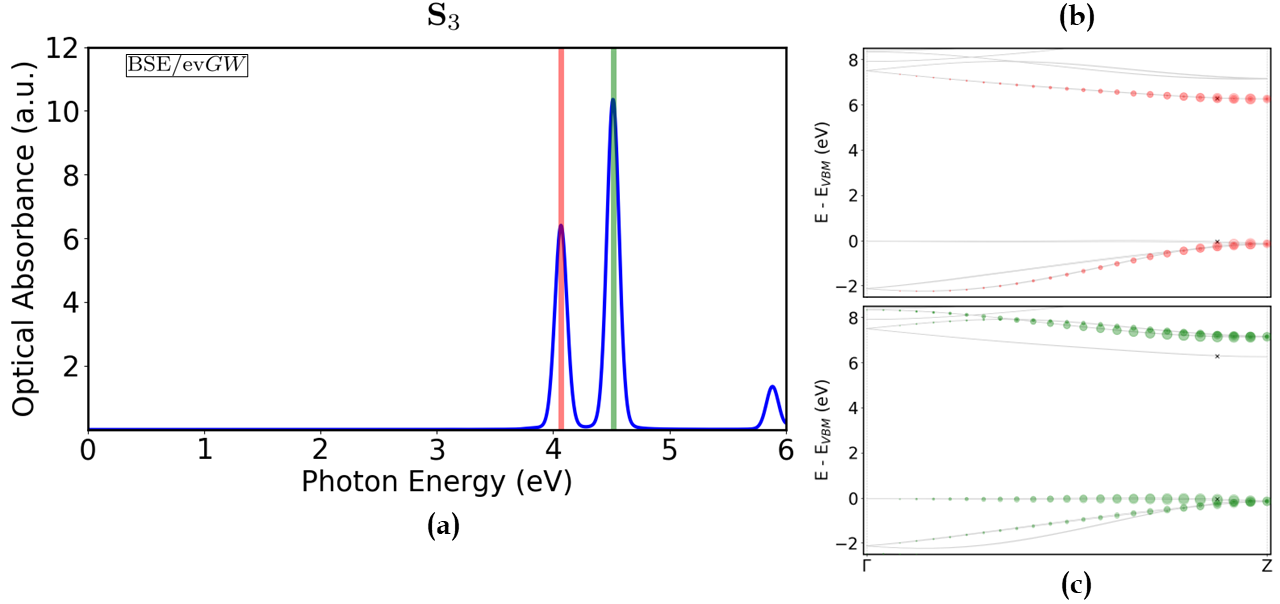} \\
    \includegraphics[scale=0.45]{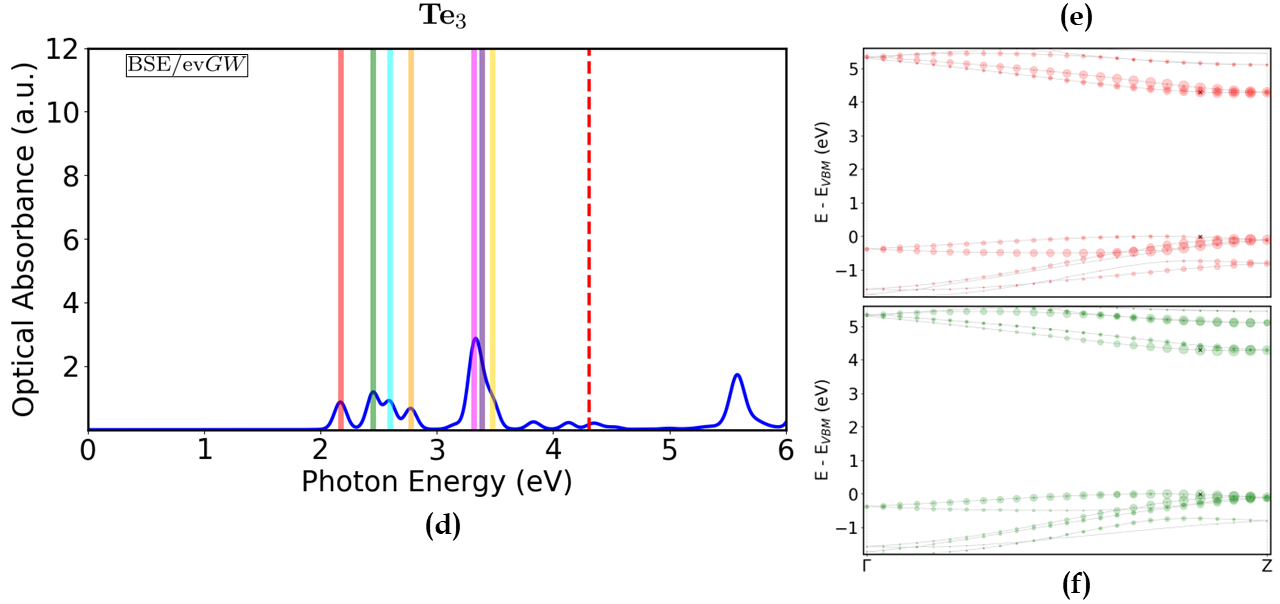}
  \end{tabular}
  \caption{[Left] Absorption spectra (solid blue) of $(\mathbf{a})$ $\mathrm{S}_3$ and $(\mathbf{d})$ $\mathrm{Te}_3$, expressed in terms of the optical absorbance $A(\omega)$, calculated at the ev$GW/$BSE level. The corresponding ev$GW-$corrected direct electronic band gaps (dashed red) are shown as a reference (6.34 eV for S$_3$). A broadening of 50 meV was used. [Right] Electronic band structures (solid grey), calculated at the ev$GW$ level, of $\mathrm{S}_3$ $(\mathbf{b}-\mathbf{c})$ and $\mathrm{Te}_3$ $(\mathbf{e}-\mathbf{f})$. The colored dots (red and green) represent the single-particle transitions contributing to the first two bright excitons, and their size is proportional to the intensity of the transition - renormalized to the highest value. The corresponding excitonic peaks are highlighted in the relative absorption spectra (solid red and green), together with other meaningful higher excitations below the electronic gap (see Fig.~12 in the SI). Energy zero is set as the top of the valence bands. SOC and semi-core corrections were included.\label{1D_GWBSE_1}}
\end{figure*}

In this section, we start analyzing the electronic band structures of the four different materials under study, first obtained at the DFT level and then refined using QP corrections within the $GW$ approximation. Then, we study their optical spectra at the highest many-body level, thus including the electron-hole (\emph{e-h}) interaction by solving the BSE.
Given their dimensionality, these materials exhibit exceptionally flat bands (especially S$_3$ and Te$_3$), which may harbor a range of exotic properties yet to be investigated, while also complicating a clear identification of the electronic band gap. In particular, these flat bands, together with the band degeneracies and the variation of the position of the band extrema in the BZ, make an effective-mass approximation near the VBM and the CBM difficult. We will return to this point in the following (Section \emph{\nameref{Exciton Model}}). \\
The QP band structures, calculated within the evGW approach, are reported in the right panels of Fig.~\ref{1D_GWBSE_1} and \ref{1D_GWBSE_2}, whereas the DFT band structures are shown in Figs. 4-5 in the Supporting Information (SI), without and with the inclusion of SOC, respectively. We refer the reader to the SI for a thorough analysis of these results. Here, we simply point out that all four wires are indirect semicondutors, with band gaps spanning from 0.4 to 2.7 eV. The corresponding values are listed in Table~\ref{tab1_1D}. \\
\noindent Concerning the QP corrections, we find that GW results are significantly affected by the strong dependence of the empty states on the cell vacuum, which hinders the convergence of most key quantities in MBPT. For further details, we refer to the SI. \\
\noindent The extremely low screening in freestanding 1D systems necessitates the use of a methodology similar to that used for systems with a 3D confinement. For instance, to address this, single-shot $G_0W_0$ calculations are followed by eigenvalue self-consistent calculations on both $G$ and $W$ (referred to as ev$GW$). The calculated values are presented in Table~\ref{tab2_1D} and in Table~1 of the SI. The lowest electronic gaps are corrected to 5.46 ($G_0W_0$) and then to 6.25 eV (ev$GW$) for S$_3$; from 3.47 ($G_0W_0$) to 4.27 eV (ev$GW$) for Te$_3$; for As$_2$S$_3$, from 2.26 ($G_0W_0$) to 3.00 eV (ev$GW$); finally, for Bi$_2$Te$_3$, from 1.54 ($G_0W_0$) to 1.65 eV (ev$GW$). Remarkably, in the latter case the electronic band gap undergoes an indirect-to-direct transition, as the QP correction at $\Gamma$ ends up to be higher than the ones at neighboring $\mathbf{k}$-points. Upon convergence, the QP gaps are increased by the ev$GW$ corrections by 3.59 (S$_3$), 2.84 (Te$_3$), 1.74 (As$_2$S$_3$), and 1.23 (Bi$_2$Te$_3$) eV according to the decreasing fundamental gap and, therefore, the increasing electronic polarizability. \\

\begin{table*}[htbp!]
\small
\renewcommand{\arraystretch}{1.5}
\caption{Calculated lowest direct electronic band gaps $(E^{\mathrm{ev}GW}_{g})$, at the $\mathrm{ev}GW$ level, together with the BSE$/\mathrm{ev}GW$ optical gaps $(E^{\mathrm{BSE}/\mathrm{ev}GW}_{opt})$ and the corresponding binding energies $(E^{\mathrm{BSE}/\mathrm{ev}GW}_{b})$ and radii $(r^{\mathrm{BSE}/\mathrm{ev}GW}_b)$ of the lowest bright excitons. In the case of an indirect QP band gap, the corresponding direct band gap is also reported in square brackets. In round brackets, the corresponding BSE$/G_0W_0$ findings can be found for comparison. The last two columns show the exciton binding energies $(E^{\mathrm{M}}_{b})$ and radii $(r^{\mathrm{M}}_b)$ obtained by the analytical model described later (Section \emph{\nameref{Exciton Model}}), using the softcore/modified softcore 1D hydrogen potential. The SOC and semi-core corrections were included. An extended table is reported in the SI.}\label{tab2_1D}
\setlength{\tabcolsep}{1pt}
\begin{tabularx}{\textwidth}{>{\centering\arraybackslash}p{1.3cm}>{\centering\arraybackslash}p{2.3cm}>{\centering\arraybackslash}p{2.9cm}>{\centering\arraybackslash}p{3.1cm}>{\centering\arraybackslash}p{2.7cm}>{\centering\arraybackslash}p{2.1cm}>{\centering\arraybackslash}p{1.6cm}}
\hline
   & $E^{\mathrm{ev}GW}_{g}$(eV) & $E^{\mathrm{BSE}/\mathrm{ev}GW}_{opt}$(eV) & $E^{\mathrm{BSE}/\mathrm{ev}GW}_{b}$(eV) & $r^{\mathrm{BSE}/\mathrm{ev}GW}_b (\si{\angstrom})$ & $E^{\mathrm{M}}_{b}$(eV) & $r^{\mathrm{M}}_b(\si{\angstrom})$ \\ \hline
    \textbf{S}$_3$ & 6.25 [6.34] & 4.07 & 2.27 (1.93) & 12.4 (12.5) & 1.80/2.19 & 6.4/6.0 \\
    \textbf{Te}$_3$ & 4.27 [4.31] & 2.17 & 2.14 (1.65) & 14.7 (14.8) & 1.62/1.94 & 7.7/7.2 \\
    \textbf{As}$_2$\textbf{S}$_3$ & 3.00 [3.04] & 2.71 & 0.33 (0.89) & 14.4 (12.00) & 1.24/1.47 & 11.2/10.5 \\
    \textbf{Bi}$_2$\textbf{Te}$_3$ & 1.65 & 0.80 & 0.85 (0.77) & 13.2 (13.2) & 0.78/0.94 & 16.7/15.5 \\
    \hline
\end{tabularx}
\end{table*}

\noindent The absorption spectra of the four 1D systems, calculated at the BSE level using ev$GW$-corrected QP states, are presented in Figs.~\ref{1D_GWBSE_1} and \ref{1D_GWBSE_2}. For comparison, the same spectra derived from $G_0W_0$-corrected states are shown in Figs. 6 and 7 of the SI. For the sake of brevity, here we focus our discussion on the ev$GW/$BSE spectra. The resulting optical band gaps are reported in Table \ref{tab2_1D}, together with the binding energies and radii of lowest bright excitons. The optical absorption is expressed here in terms of the frequency-dependent optical absorbance (or absorption coefficient) $A(\omega) = (\omega/c) \, \mathrm{Im}[\alpha_{\mathrm{1D}}(\omega)]$, where we have defined the 1D macroscopic electronic polarizability of the wire as $\alpha_{\mathrm{1D}}(\omega) = (\varepsilon(\omega) -1) \, \mathrm{S}/(4\pi)$, with $\varepsilon(\omega)$ the dielectric function and $\mathrm{S}$ the surface area of the supercell transversal to the axis of the wire, thus making the intensity of the spectra independent of the supercell vacuum. \\

\begin{figure*}[h!]
  \centering
  \begin{tabular}{@{}c@{}}
    \includegraphics[scale=0.4]{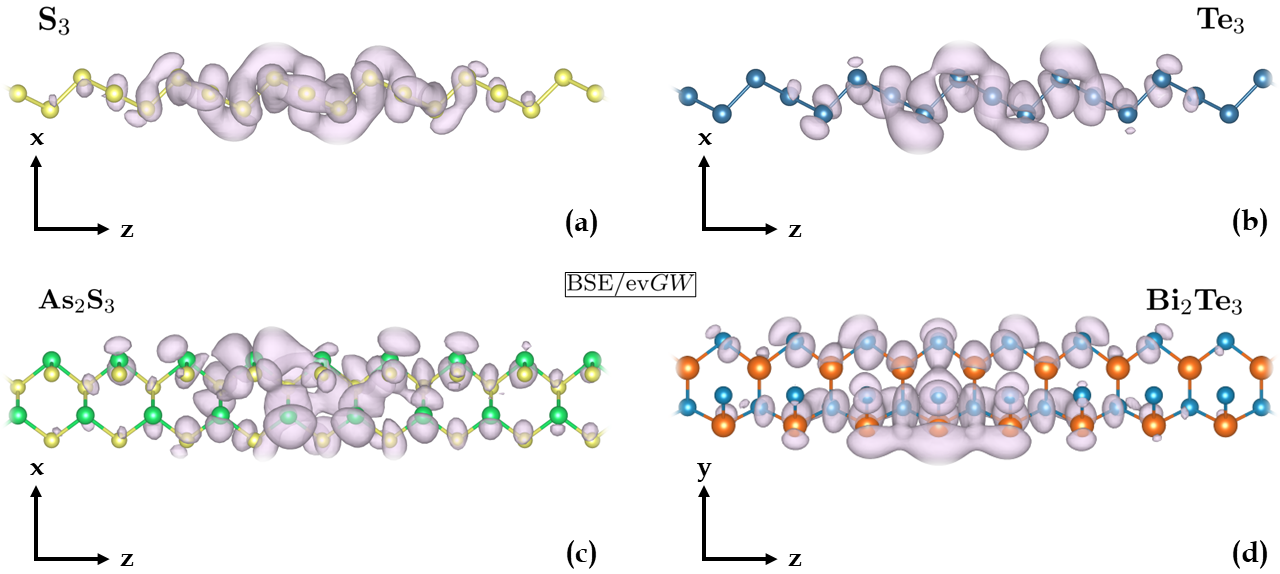} \\[\abovecaptionskip]
  \end{tabular}
  \caption{Plots in direct space of the excitonic wavefunctions of the first bright exciton of $(\mathbf{a})$ $\mathrm{S}_3$, $(\mathbf{b})$ $\mathrm{Te}_3$, $(\mathbf{c})$ $\mathrm{As}_2\mathrm{S}_3$ and $(\mathbf{d})$ $\mathrm{Bi}_2\mathrm{Te}_3$, calculated at the ev$GW/$BSE level. The position of the hole was chosen based on the localization of the valence electrons contributing to the exciton. The number of cell repetitions in the periodic direction was increased until the wavefunction decayed to zero.\label{1D_EXC_evgw}}
\end{figure*}

\begin{figure*}[h!]
  \centering
  \begin{tabular}{cc}
    \includegraphics[scale=0.45]{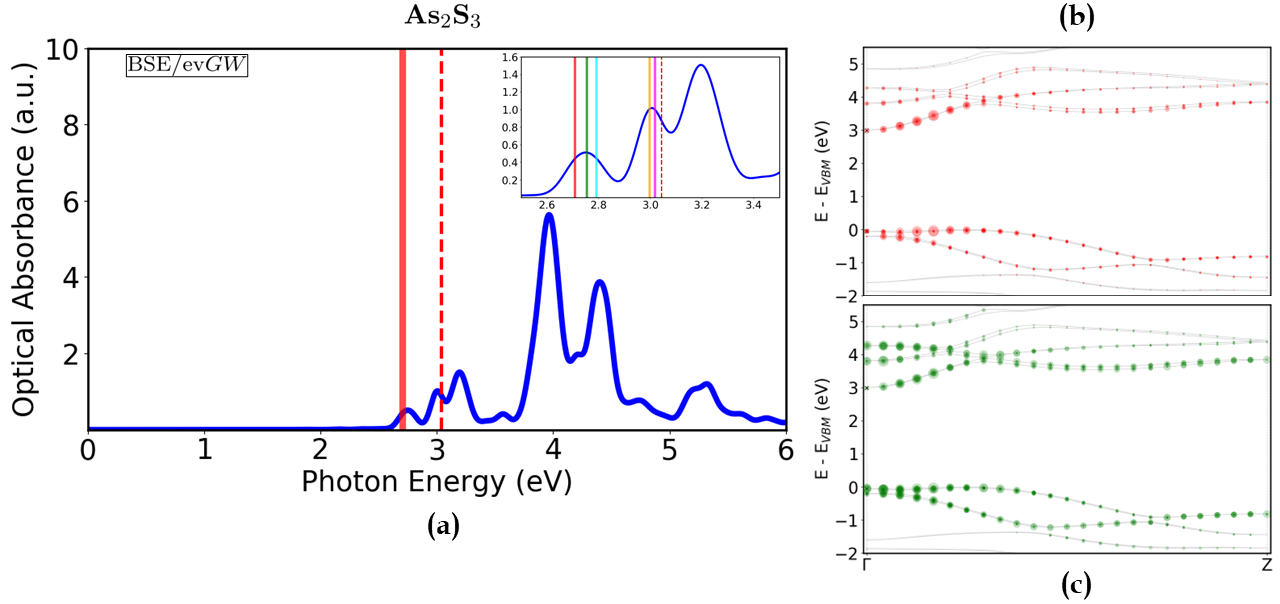} \\
    \includegraphics[scale=0.45]{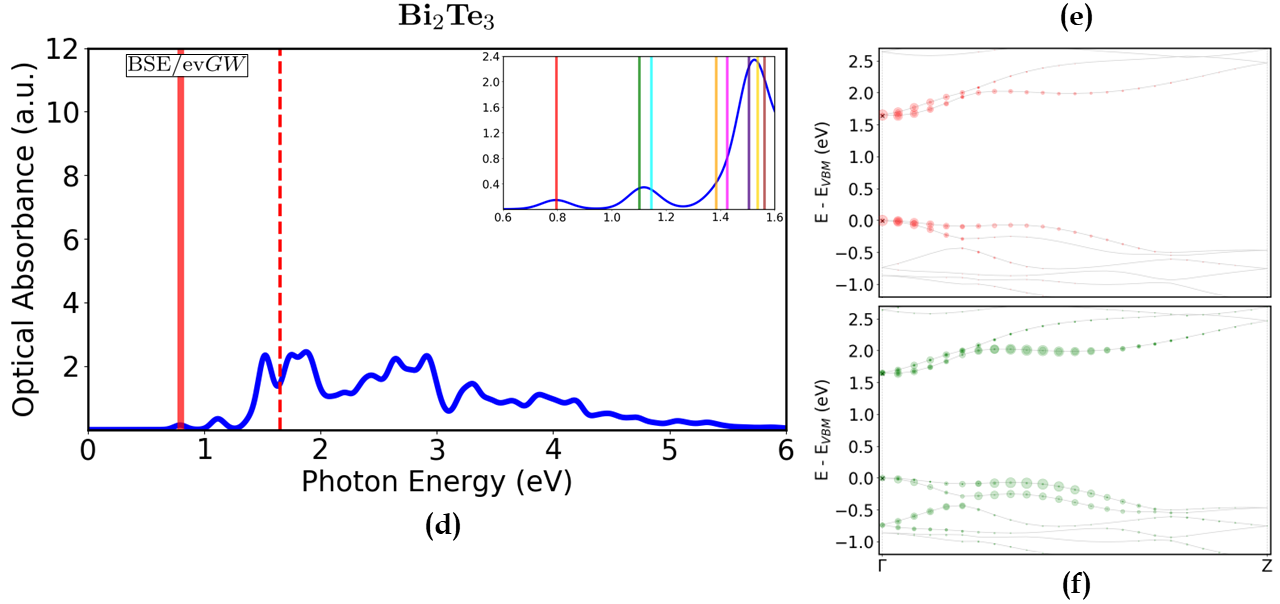}
  \end{tabular}
  \caption{[Left] Absorption spectra (solid blue) of $(\mathbf{a})$ $\mathrm{As}_2\mathrm{S}_3$ and $(\mathbf{d})$ $\mathrm{Bi}_2\mathrm{Te}_3$, expressed in terms of the optical absorbance $A(\omega)$, calculated at the ev$GW/$BSE level. The corresponding ev$GW-$corrected direct electronic band gaps (dashed red) are shown as a reference. A broadening of 50 meV was used. [Right] Electronic band structures (solid grey), calculated at the ev$GW$ level, of $\mathrm{As}_2\mathrm{S}_3$ $(\mathbf{b}-\mathbf{c})$ and $\mathrm{Bi}_2\mathrm{Te}_3$ $(\mathbf{e}-\mathbf{f})$. The colored dots (red and green) represent the single-particle transitions contributing to the first two bright excitons, and their size is proportional to the intensity of the transition - renormalized to the highest value. The corresponding excitonic peaks are highlighted in the relative absorption spectra (solid red and green) and in their insets, together with other meaningful higher bound states below the electronic gap (see Figs.~13-14 in the SI). Energy zero is set as the top of the valence bands. SOC and semi-core corrections were included.\label{1D_GWBSE_2}}
\end{figure*}

\noindent The analysis of the results obtained from the BSE calculations is divided in two parts, based on the common features of the four materials considered. We first investigate the two elemental wires S$_3$ and Te$_3$. As shown in Fig.~\ref{1D_GWBSE_1}, the spectra of these two systems, below and around the direct QP gap, consist of relatively highly-intense optical excitations that are well-separated in energy, resembling what one might expect from strongly confined systems. Below the gap, they represent dipole-allowed excitons, neither whose intensities nor excitation energies follow a simple law in the main quantum number. With an optical gap of 4.07 (S$_3$) and 2.17 (Te$_3$) eV, the lowest bright excitations are located in the near UV and in the visible range, respectively. The lowest exciton bound states are much below the QP gaps in Table~\ref{tab2_1D}. Consequently, they exhibit a significant binding energy above 2 eV.
In the case of S$_3$ (Fig.~\ref{1D_GWBSE_1}a), two prominent excitations appear below the direct electronic band gap (6.34 eV), followed by a smaller peak. They are bound excitonic states with varying binding energy.
The first exciton, which marks the onset of optical absorption, occurs at 4.07 eV, with a binding energy of 2.27 eV. As shown in Fig.~\ref{1D_GWBSE_1}b, it arises from transitions involving the third highest degenerate occupied band and the first (degenerate) conduction band, close to the Z point. The corresponding excitonic wavefunction, whose radius (\emph{i.e.}, their lateral extent), is estimated to be \SI{12.4}{\angstrom}, is depicted in Fig.~\ref{1D_EXC_evgw}a.
The second and most intense peak (4.51 eV) instead stems from transitions involving mainly the doubly-degenerate valence band and the doubly degenerate second conduction band, again close to the Z point, as illustrated in Fig.~\ref{1D_GWBSE_1}c. \\

\noindent In the case of Te$_3$ (Fig.~\ref{1D_GWBSE_1}d), the optical absorption spectrum consists of several bright excitations below the direct electronic band gap (4.31 eV). Tellurium has been broadly investigated in the past years. Particularly, 2D tellurium --- known as tellurene ---, found in different allotropic forms \cite{grillo2022}, has emerged as a promising semiconductor of increasing with a thickness-tunable band gap \cite{huang2017epitaxial}, relatively high carrier mobility \cite{tong2020stable, amani2018solution} and remarkable electronic and optical characteristics, with optical band gaps ranging from 0.84 to 1.46 eV and a light absorption as high as 50$\%$ \cite{grillotellurene}. \\
In its 1D counterpart, the first bright exciton appears at 2.17 eV, thus with a binding energy of 2.14 eV. It is noteworthy to mention that previous calculations done at the $G_0W_0/$BSE level predict for Te$_3$ a similar exciton binding energy of 2.07 eV \cite{pan2018dependence} and 2.35 eV \cite{andharia2018exfoliation}, with a $G_0W_0$ direct gap of, respectively, 4.23 and 4.44$-$4.59 eV, surprisingly close to our ev$GW$ estimated direct gap of 4.31 eV. The apparent agreement in the gap is related to the missed inclusion of SOC in both the aforementioned calculations, leading to an overestimation of the QP corrections. The exciton binding energy, however, is practically not affected by the SOC. 
The transitions contributing to this peak involve the six uppermost valence bands, and the three lowest conduction bands, with major contributions coming from the second highest valence band and the second lowest conduction band, spanning the entire BZ (Fig.~\ref{1D_GWBSE_1}e).
The corresponding excitonic wavefunction, with a Bohr radius of \SI{14.7}{\angstrom}, is depicted in Fig.~\ref{1D_EXC_evgw}b, closely resembling that of S$_3$. The second bright exciton, occuring at 2.45 eV, involves the same bands present in the previous one, with a slightly different distribution of the single-particle contributions (Fig.~\ref{1D_GWBSE_1}f). Other meaningful higher bound states, highlighted in Fig.~\ref{1D_GWBSE_1}d by different colors, are found below the electronic gap. The contributions to these peaks are illustrated in more details in Fig.~12 in the SI and will not be discussed here. \\

\noindent In contrast to the elemental cases, the absorption spectra of As$_2$S$_3$ and Bi$_2$Te$_3$ notably resemble a continuum of transitions, similar to those observed in 3D and 2D materials, which give rise to a continuous variation with some spectral maxima compared to the elemental chalcogen chains. In particular, bulk Bi$_2$Te$_3$ has recently gained a wide interest because of its topological features \cite{chen2009experimental} and its successful exfoliation \cite{goyal2010mechanically}. \\
\noindent These materials exhibit lower QP band gaps, exciton binding energies ($< 1$ eV) and peak intensities. In the As$_2$S$_3$ spectrum, two main peaks, composed of several excitations, appear below the direct electronic gap (3.04 eV), as shown in Fig.~\ref{1D_GWBSE_2}a and highlighted in the related inset. The first of these peaks is composed of three bright excitations, occurring at 2.71, 2.75 and 2.79 eV, falling within the blue-violet region of the visible spectrum and exhibiting small binding energies of 0.33, 0.29 and 0.25 eV, respectively, if measured with respect to the QP gap. Being so close in energy, these excitons originate from the same single-particle transitions, mainly involving the first two degenerate occupied bands and the first six degenerate unoccupied bands, concentrated around $\Gamma$. These band pair contributions are shown in detail in Fig.~\ref{1D_GWBSE_2}b-c and in Fig.~13a in the SI. The excitonic wavefunction of the lowest bound state is depicted in Fig.~\ref{1D_EXC_evgw}c and presents an estimated radius of \SI{14.4}{\angstrom}. Similarly, the second more intense peak arises from two nearly-degenerate excitations occurring at 3.00 and 3.02 eV. As shown in Fig.~13b-c in the SI, they stem from transitions between the first two degenerate occupied bands and the first and the fourth degenerate unoccupied bands, concentrated around $\Gamma$. To conclude the discussion, it is worth to mention that, in the case of As$_2$S$_3$, the ev$GW$ calculations dramatically modify the excitonic spectrum. In particular, the self-consistent procedure causes the contributions to the excitons to migrate from the region around Z (see Fig.~7a-c in the SI) to that around $\Gamma$, as previously shown. The main effect of this redistribution is a drastic and anomalous drop of the exciton binding energy of the lowest bright exciton from 0.89 to 0.33 eV, with related increase of the exciton radius from 12.0 to 14.4 \si{\angstrom}. \\
\noindent Finally, in the Bi$_2$Te$_3$ case (Fig.~\ref{1D_GWBSE_2}d), the first exciton peak below the gap (1.65 eV) occurs at 0.8 eV, thus in the near-IR region. 
This gives an exciton binding energy of 0.85 eV, which is again significantly lower than that of the other three materials. The corresponding excitonic wavefunction (Fig.~\ref{1D_EXC_evgw}d) has a radius of \SI{13.2}{\angstrom}. This first excitation, as well as the others with higher energies below the gap, mainly arises from transitions involving the two highest degenerate valence bands and the two lowest degenerate conduction bands close to $\Gamma$ (see Fig.~\ref{1D_GWBSE_2}e). The second peak, at 1.1 eV, is composed of two excitations, whose transitions stem from the same bands (see Fig.~\ref{1D_GWBSE_2}f and Fig.~14a in the SI) in the region between $\Gamma$ and $1/2 \, \Gamma Z$. These first two peaks have a very low intensity. The last peak below the gap is indeed the most intense and it is composed of three nearly-degenerate excitations (see inset in Fig.~\ref{1D_GWBSE_2}d), located at 1.51, 1.54 and 1.56 eV. These excitons are composed by a bigger number of bands and a bigger energy range, as shown in Fig.~14e-g in the SI. \\
\noindent In summary, in the absorption spectra of the four 1D materials, enhanced excitonic effects are visible, in particular, strongly bright bound excitons. For the freestanding quantum wires under consideration, the low screening, due to the 1D electronic system, plays a prominent role. Due to the reduced dimensionality, electrons and holes experience stronger Coulomb interactions, as screening is less effective in 1D compared to bulk systems. This leads to the formation of tightly bound excitons with large binding energies, significantly influencing the optical properties. \\

\subsection{Understanding 1D excitons}\label{Exciton Model}

As shown in Table~\ref{tab2_1D}, the lowest-energy excitons exhibit large binding energies on the order of $1-2$ eV (except for the case of As$_2$S$_3$). These strong excitonic effects can be understood through their electronic structures (Figs.~\ref{1D_GWBSE_1} and \ref{1D_GWBSE_2}) and the resulting electronic screening. \\
\noindent Since the two-particle states primarily arise from the lowest conduction and highest valence bands near Z for S$_3$ and Te$_3$ (or near $\Gamma$ for As$_2$S$_3$ and Bi$_2$Te$_3$), we employ a two-band model with the effective-mass approximation (EMA) centered around these high-symmetry points. To this end, we make use of the electronic band structures calculated at the DFT level (reported in Fig.~5 in the SI) with SOC included to extract effective masses of the electrons and holes forming the lowest exciton bound state. \\
\noindent The internal motion of the bound electron-hole pair along the wire axis $z$ is described, within the EMA, by the Schrödinger equation:

\begin{equation}
\left\{ -\frac{\hbar^2}{2\mu} \frac{d^2}{dz^2} + W(z) \right\} \phi(z) = - E_b^M \phi(z),
\label{eq:schrodinger}
\end{equation}

\vspace{0.25cm}

\noindent where $W(z)$ represents the attractive screened Coulomb interaction. The wavefunction $\phi(z)$ describes the internal motion of the lowest exciton with binding energy $E_b^M$. The kinetic energy is ruled by $\mu$, i.e. the reduced effective mass of the electron-hole pair, taking values $\mu = 0.39$, $0.30$, $0.16$, and $0.10$, in units of the electron mass, for S$_3$, Te$_3$, As$_2$S$_3$, and Bi$_2$Te$_3$, respectively. \\
\noindent The screened potential $W(z)$ is derived from the Poisson equation, modeling the quantum wire as a cylinder with radius $R$ and a static 1D electronic polarizability $\alpha^0_{\text{1D}}=S(\epsilon_1(0)-1)/4\pi$ calculated within the independent particle picture using  DFT eigenvalues and eigenfunctions, in a similar fashion as in the 2D case \cite{bechstedt2021beyond}. Here, $\epsilon_1(0)$ is the real part of the supercell dielectric function at $\omega=0$ for light polarized in the $z$ direction, and S is the xy area of the supercell. \\We obtain the polarizabilities $\alpha^0_{\text{1D}} =$ 12.06 (S$_3$), 17.43 (Te$_3$), 26.05 (As$_2$S$_3$), and 51.63 (Bi$_2$Te$_3$) \AA$^2$. The Fourier transform of $W(z)$ is given by \cite{bechstedt2025}:

\begin{equation}
\tilde{W}(q) = -e^2 \frac{\tilde{V}_{\text{bare}}(q)}{1 + \alpha^0_{\text{1D}} \, q^2 \, \tilde{V}_{\text{bare}}(q)},
\label{eq:fourier}
\end{equation}

\vspace{0.25cm}

\noindent where $\tilde{V}_{\text{bare}}(q)$ is the Fourier-transformed bare Coulomb potential, averaged over the wire cross-section using different distribution functions $f(\rho)$. Because of the long-range interaction between charged particles, only small wave vectors $q$ play a role. Therefore, a possible modification of the denominator by the non-locality of the screening reaction is omitted. We assume a screening reaction which is localized at the rod surface with $f(\rho) = \frac{1}{2 \pi R} \delta(\rho - R)$ or homogeneously distributed over the wire cross-section with $f(\rho) = \frac{1}{\pi R^2} \theta(R - \rho)$. The resulting bare potentials are

\begin{align}
\tilde{V}_{\text{bare}}(q) &= 2K_0(|q|R), \\
\tilde{V}_{\text{bare}}(q) &= \frac{4}{q^2 R^2} [1 - |q| \, R \, K_1(|q|R)],
\end{align}

\vspace{0.25cm}

\noindent with zeroth- ($K_0$) and first-order ($K_1$) modified Bessel functions. In real space, they become, respectively, the softcore and modified softcore 1D Coulomb potentials \cite{loudon2016one}

\begin{align}
V_{\text{bare}}(z) &= \frac{1}{\sqrt{z^2 + R^2}}, \\
V_{\text{bare}}(z) &= \frac{2}{R^2} [\sqrt{R^2 + z^2} - |z|].
\end{align}

\vspace{0.25cm}

\noindent The complexity of the screened potential (\ref{eq:fourier}) forbids an analytic solution of the exciton problem (\ref{eq:schrodinger}). Therefore, we apply a variational treatment with the trial function:

\begin{equation}
\phi(z) = \sqrt{\frac{2s^3}{a^3_{\text{exc}}}} |z| e^{-s|z|/a_{\text{exc}}},
\label{eq:trial}
\end{equation}

\vspace{0.25cm}

\noindent where $s$ is a variational parameter and $a_{\text{exc}} = a_B \, m_e/\mu$ is the effective Bohr radius. In the limit $s=1$, (\ref{eq:trial}) represents the ground-state wavefunction of the 1D hydrogen atom with the bare potential $W(z) = -\frac{e^2}{|z|}$ \cite{loudon1959one}. From (\ref{eq:schrodinger}), (\ref{eq:fourier}), and (\ref{eq:trial}), we obtain the variational expression:

\begin{equation}
E^M_b(s) = R_{\text{exc}} \left[ -s^2 + s \frac{4}{\pi} \int_{0}^{\infty} dt \frac{1 - 3t^2}{(1 + t^2)^3} \left( -\frac{1}{e^2} \right) \, \tilde{W} \left( \frac{2s}{a_{\text{exc}}} t \right) \right].
\label{eq:variational}
\end{equation}

\vspace{0.25cm}

\noindent with $R_{\text{exc}} = R_H \, \mu/m_e$ and R$_H$=1Ry. Its maximum at $s=s_0$, for which $\frac{d}{ds}E^M_b(s)|_{s=s_0}=0$, yields the exciton binding energies $E_b^M$ and exciton radii $r_b^M$, listed in Table~\ref{tab2_1D}. \\
\noindent The estimated binding energies vary with the softcore or modified softcore 1D hydrogen potential used by less than 20\%. This effect is reduced for the average electron-hole distances $r_b^M$. Consequently, one can conclude that the distribution of interaction of the charged particles over the wire cross-section is of minor influence. More important is the screening itself. Despite the small wire radius $R$, the screening significantly reduces the value $E_b^M$ with respect to $R_{\text{exc}}$ by a factor of $2-3$. In the case of the excitonic radii $r_b^M$, the screening effect is somewhat enhanced. The screening reaction of the electron gas in the considered 1D systems is of extreme importance to understand the electron-hole attraction. Indeed, the reaction can be described by a 1D electronic polarizability as indicated in formula (\ref{eq:fourier}). \\
\noindent The comparison of the model estimates $E_b^M$ and $r_b^M$ with the corresponding ab initio values $E_b^{\text{BSE/evGW}}$ (or $E_b^{\text{BSE/G}_0\text{W}_0}$) and $r_b^{\text{BSE/evGW}}$ (or $r_b^{\text{BSE/G}_0\text{W}_0}$), reported in Table~\ref{tab2_1D}, shows the same chemical trends along the row S$_3$, Te$_3$, As$_2$S$_3$, and Bi$_2$Te$_3$, but also very similar values. This holds especially for the monoatomic chains S$_3$ and Te$_3$, where the binding energies agree very well, applying the correspondence BSE/G$_0$W$_0$ $\leftrightarrow$ \textit{softcore potential} and BSE/evGW $\leftrightarrow$ \textit{modified softcore potential}. In the case of the more complex systems As$_2$S$_3$ and Bi$_2$Te$_3$, the quantitative agreement is reduced, but the same order of magnitude is still guaranteed. In any case, we confirm that the exciton binding can be reliably predicted by the model based on a parabolic two-band model, generalized 1D hydrogen Coulomb potentials, and a screening reaction of the 1D electron gas described by means of a static electronic polarizability.

\section{Conclusions}

We have presented a first-principles investigation of a novel class of potentially exfoliable atomic wires.
We explored the structural, energetic, vibrational, electronic and optical properties of four promising chain-like materials --- S$_3$, Te$_3$, As$_2$S$_3$, and Bi$_2$Te$_3$ --- using DFT, DFPT and MBPT, including $GW$ corrections and BSE calculations. \\
\noindent The selected chain systems are dynamically stable, with the exception of As$_2$S$_3$, where the lowest TA phonon branch exhibits a slight negative dispersion, suggesting a tendency to relax into a larger unit cell at 0 K.
Additionally, the phonon dispersions display the characteristic twisting acoustic mode of realistic one-dimensional materials.
Interestingly, the IR spectra of the homopolar S$_3$ and Te$_3$ chains exhibit one weak absorption peak where the dipole moment arises from the geometrical conformation rather than from differences in electronegativity.
In contrast, As$_2$S$_3$ and Bi$_2$Te$_3$, with partly ionic bonding, display, as expected, more IR-active phonons and significantly stronger absorption peaks.
\noindent The chain materials exhibit unique band structures with flat bands and substantial band gap separations, indicating potential for exotic electronic phenomena related to electrons and/or holes with relatively heavy effective masses. Because of the weak electronic screening, the band structures are significantly influenced by single-particle QP effects, at least in Hedin's $GW$ approximation. The gaps and interband distances are greatly increased compared to the KS values from the DFT framework. The gaps are opened by $1-2.8$ eV in the first $G_0W_0$ iteration step. Further openings have been observed in the limit of converged $GW$ calculations, at least with respect to the single-particle eigenvalues. Interestingly, the QP effects also influence the band dispersion. While the KS band structures indicate indirect semiconductors with the maxima of the uppermost valence band on the $\Gamma Z$ line, the self-consistent QP treatment tends to indirect-to-direct transitions with direct gap at $Z$ in the homopolar S$_3$ and Te$_3$ chains and at $\Gamma$ in the polar 1D materials As$_2$S$_3$ and Bi$_2$Te$_3$. \\
\noindent The most relevant many-body features appear in the optical spectra. Besides the QP blueshift of the absorption threshold, a significant redistribution of the spectral strengths occurs due to the electron-hole attraction, the excitonic effects. Peaks of bound excitons appear below the single-particle absorption edge, defined by the QP gap. The homopolar chains S$_3$ and Te$_3$ show significant exciton binding energies, indicative of highly localized excitonic states, while the ionic compounds As$_2$S$_3$ and Bi$_2$Te$_3$ exhibit a continuum of transitions with smaller binding energies. \\
\noindent The extremely strong excitonic effects are explained introducing a novel model based on the effective mass approximation of the kinetic energy of the internal electron-hole motion and a screened Coulomb potential in 1D systems. The latter describes the screening by a 1D static electronic polarizability of the electron gaps in the wires. Indeed, the large exciton binding energies, of the order of $0.8-2.3$ eV, can be explained by the small electronic polarizabilities and the large interband masses of $0.1-0.7$ m$_e$, at least for freestanding chains. \\
\noindent Given these extreme properties, S$_3$ and Te$_3$ are promising candidates for optoelectronic applications, particularly in UV- and visible-absorption devices due to their strong excitonic effects. In the case of the polar chain materials, additional strong absorption appears in the range of optical phonons. The lower exciton binding energies and broader absorption spectra of As$_2$S$_3$ and Bi$_2$Te$_3$ make them suitable for photovoltaic and infrared applications. These findings suggest that exfoliable 1D wires could be key materials for future nanoscale electronic and optoelectronic technologies. \\

\section{Methods}

\noindent The calculation of ground-state properties within the framework of DFT is carried out using the Quantum ESPRESSO (QE) distribution \cite{giannozzi2009quantum, giannozzi2017advanced}.
Phonon dispersions are computed using density-functional perturbation theory (DFPT) \cite{baroni2001phonons}, using $\mathbf{q}$-grids of $1 \times 1 \times 6$ (Te$_3$), $1 \times 1 \times 10$ (S$_3$ and Bi$_2$Te$_3$) and $1 \times 1 \times 20$ (As$_2$S$_3$). Pseudopotentials are taken from the SSSP library \cite{prandini2018precision}, v1.1 PBE efficiency, using the suggested kinetic energy cut-offs, and a minimum vacuum distance of \SI{21}{\angstrom} (up to a maximum of \SI{27}{\angstrom}). The correct long-wavelength behavior of the acoustic phonons is imposed using the acoustic sum rule that includes rotational symmetries, implemented as described in Ref. \citenum{lin2022general}.
The IR spectra are obtained from the \verb|dynmat| post-processing module of QE, starting from the eigenmodes and Born effective charges previously computed. The intensities of the peaks are smeared with a Gaussian function of half-width 9 cm$^{-1}$, and normalized by the volume of the wires ($4\pi/V_0$), which we define as the quantum volume \cite{cococcioni2005electronic} calculated using the Quantum ENVIRON package \cite{andreussi2012revised}. \\
Concerning electronic properties, a norm-conserving, fully-relativistic pseudopotential from the PseudoDojo repository (v0.4) \cite{PseudoDojo} is employed to account for spin-orbit coupling (SOC) and include semi-core electrons, using an exchange-correlation functional within the generalized gradient approximation (GGA) according to Perdew, Burke and Ernzerhof (PBE) \cite{PBE}.
For these, at convergence, kinetic energy cut-offs of 110 (S$_3$), 70 (Te$_3$), 80 (As$_2$S$_3$) and 100 (Bi$_2$Te$_3$) Ry are chosen. A uniform Monkhorst-Pack \cite{monkhorst1976special} $\mathbf{k}$-point mesh, with dimension of $1\times1\times12$, is employed.
To prevent interaction between periodic replicas, a minimum vacuum region of \SI{16}{\angstrom} along the non-periodic ($xy$) directions is introduced. Structural relaxation are considered as converged when the maximum component of the residual ionic forces dropped below $10^{-8}$ Ry/Bohr. \\
From the DFT eigenvalues and eigenvectors obtained, MBPT calculations are carried out using the Yambo code \cite{marini2009yambo, sangalli2019many}, specifically employing the $G_0W_0$ and eigenvalue self-consistent $GW$ (ev$GW$) methods for the quasi-particle (QP) corrections of the electronic states and the BSE to account for the \emph{e-h} interaction in the excited states, e.g., in the optical absorption spectra \cite{BSE1,BSE2,BSE3,BSE4}. For the $GW$ calculations, cut-offs of 47899, 77927, 59085, and 47115 $\mathbf{G}$-vectors are used for the exchange part of the electron self-energy $\Sigma_x$, respectively for S$_3$, Te$_3$, As$_2$S$_3$ and Bi$_2$Te$_3$, while 12 Ry are used for the correlation part of the self-energy $\Sigma_c$.
Additionally, 1582 (S$_3$), 1402 (Te$_3$), 1452 (As$_2$S$_3$) and 622 (Bi$_2$Te$_3$) empty bands are included in the calculation of the $\Sigma_c$ \footnote{Convergence with respect to empty states has been proven to be a critical factor in 1D systems. For further details, we refer the reader to the SI.}. 
A cylindrical cut-off to the Coulomb potential along the non-periodic directions ($xy$) is also used, as implemented in the Yambo code \cite{sangalli2019many}, to avoid artificial interaction between a wire and its replica. The $GW$ calculations are performed using the plasmon-pole approximation (PPA) to model the frequency dependence of the dielectric function, as implemented in the Yambo code.
For the BSE Hamiltonian, a total of 10 occupied states and 10 unoccupied states are employed. 
The convergence with respect to the $\mathbf{k}$-points for both the QP corrections and the BSE is carefully checked, a $\mathbf{k}$-point grid of $1 \times 1 \times 48$ has been used. \\


\section*{Supporting Information Available}
Discussion on the dependence of the empty states on the cell vacuum and results for the prototypical case of MgN$_2$. Electronic band structuresat DFT-PBE level of theory for the four nanowires, with and without SOC. Absorption spectra and electronic band stuctures at G$_0$W$_0$/BSE level and electronic band stuctures at evGW level. Detailed table summarizing the MBPT results.

\noindent \textbf{Conflict of Interest}: The authors declare no competing financial interest

\section*{Acknowledgments}

S.G., M.P. and O.P. acknowledge PHOTO, CN1 Spoke6, TIME2QUEST, ECoE, CINECA.
S.G., C.C. and N.M. acknowledge NCCR MARVEL, a National Centre of Competence in Research, funded by the Swiss National Science Foundation (grant number 205602). 
C.C. thanks L. Bastonero, N. Rivano and C. Lin for useful discussions
\bibliography{bibliography}
\end{document}


\maketitle
The Supporting Information contains:
\begin{itemize}
    \item discussion of the dependence of the empty states on the cell vaccum, and results for the prototypical case of MgN$_2$ (Section \emph{\nameref{MgN2}}).
    \item the electronic band strucutres at DFT-PBE level of theory, with and without SOC (Section \emph{\nameref{DFTPBE_band}});
    \item supporting analysis for the four wires at MBPT level of theory: \emph{\nameref{app:g0w0}} and \emph{\nameref{app:evgw}}. We provide in Table \ref{tab3_1D} a detailed summary of the results, at both G$_0$W$_0$ and evGW level.
\end{itemize}

\section{Dependence of the empty states on the cell vacuum}\label{MgN2}
\subsection{The prototypical case of MgN$_2$}
\begin{figure*}[h!]
  \centering
  \begin{tabular}{@{}c@{}}
    \includegraphics[scale=0.4]{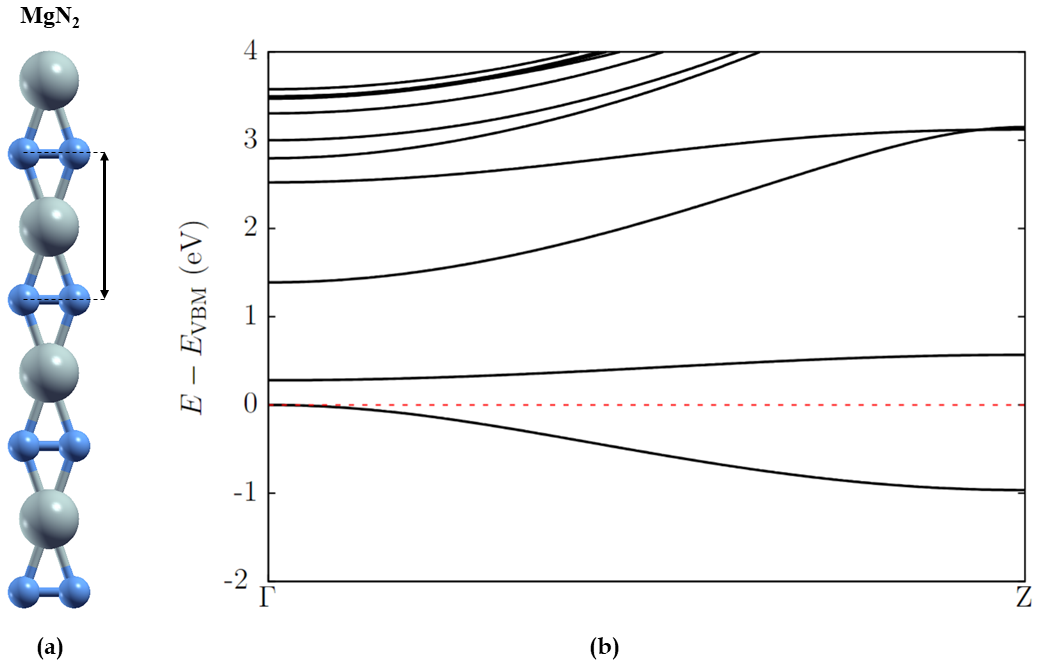} \\[\abovecaptionskip]
  \end{tabular}
  \caption{Optimized geometric structure $(\mathbf{a})$ and electronic band structure $(\mathbf{b})$ of $\mathrm{MgN}_2$. Credits to Ref. \citenum{cignarella2024searching} for the structures provided. Energy zero is set as the top of the valence bands, highlighted with a red dotted line.\label{mgn2_1}}
\end{figure*}

\begin{figure*}[h!]
  \centering
  \begin{tabular}{@{}c@{}}
    \includegraphics[scale=0.2]{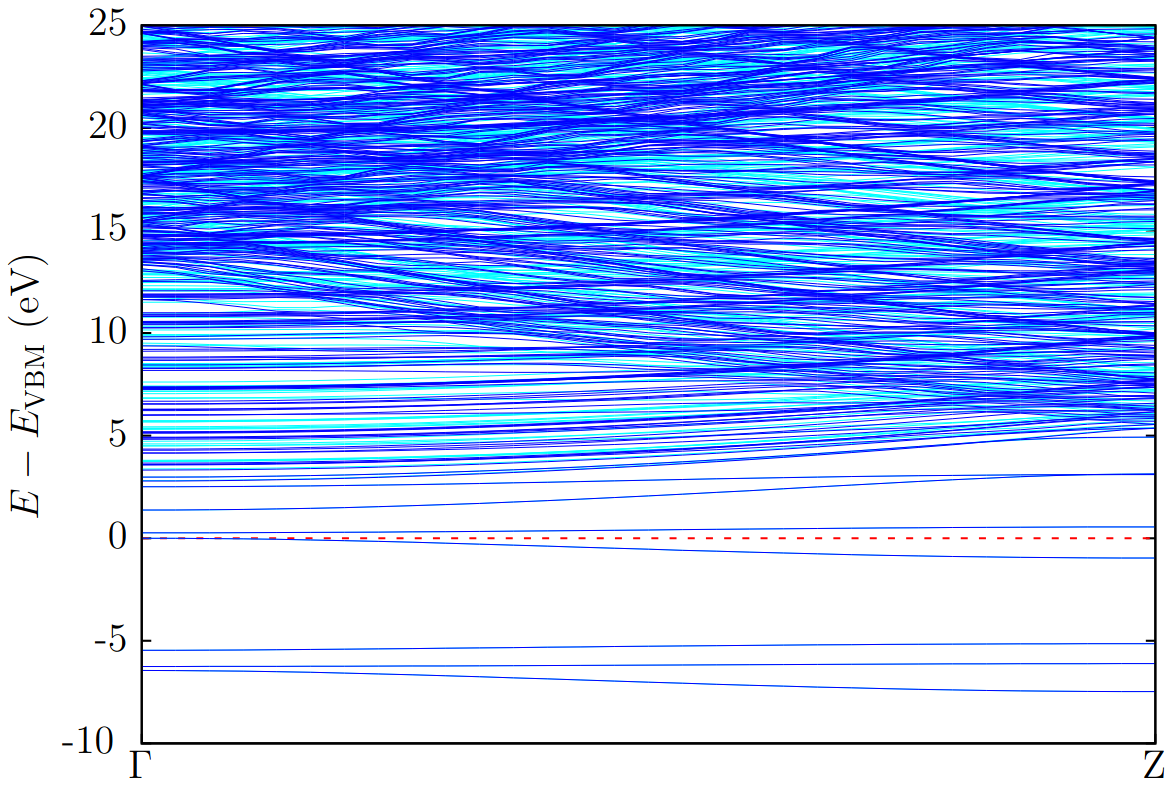} \\[\abovecaptionskip]
  \end{tabular}
  \caption{Calculated band structure, at the DFT level, of the MgN$_2$ wire using a vacuum of \SI{16}{\angstrom} (cyan) and \SI{18}{\angstrom} (blue) in the non-periodic directions. The electronic bands are well converged only up to few eV above the Fermi level (red dashed line). Energy zero is set as the top of the valence bands.\label{VACUUM_1}}
\end{figure*}

\begin{figure*}[h!]
  \centering
  \begin{tabular}{@{}c@{}}
    \includegraphics[scale=0.2]{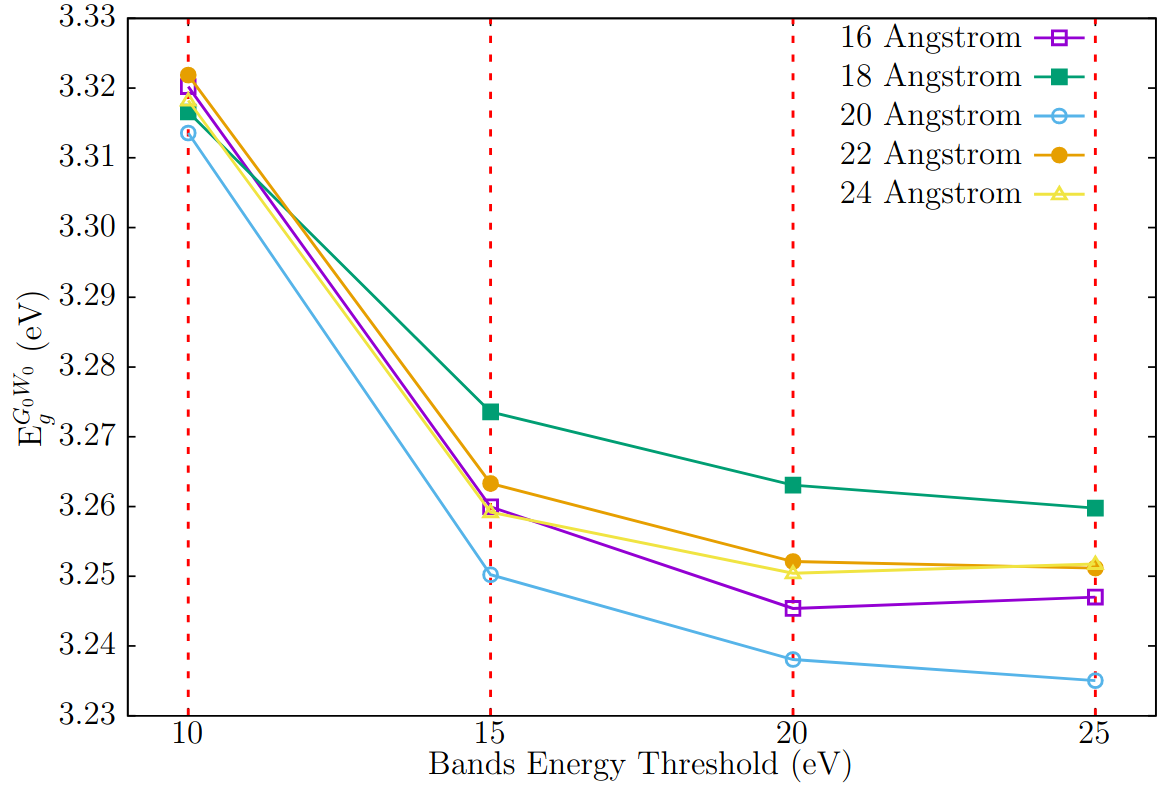} \\[\abovecaptionskip]
  \end{tabular}
  \caption{Convergence of the $G_0W_0$-corrected band gap of MgN$_2$ with respect to the energy threshold for empty bands included in the summations. The different curves represent the different vacuum used in the calculations. A vacuum of 15–16 \si{\angstrom} and an energy threshold of $20$ eV (from the Fermi level) are generally sufficient to converge the correction to the band gap, resulting in a discrepancy of less than $1\%$ compared to a vacuum of up to \SI{30}{\angstrom}.\label{VACUUM_2}}
\end{figure*}

\noindent DFT calculations of periodic low-dimensional materials such as nanowires, 2D materials, and thin films, require the inclusion of a vacuum region for achieving accurate results. These materials are often simulated using periodic boundary conditions, which replicate the unit cell throughout space. For low-dimensional systems, these periodic replicas can cause unintended interactions between the material and its periodic images in the non-periodic directions. By including a sufficiently large vacuum region around the material, typically in the range of 15-20 \si{\angstrom}, one tries to minimize these spurious interactions, particularly vdW forces and electrostatic interactions. The inclusion of vacuum is therefore a fundamental step in DFT simulations of low-dimensional materials to ensure the physical relevance and reliability of the calculated properties. \\
\noindent In recent years, researchers have been developing advanced techniques within DFT to address the limitations of the traditional approach of including large vacuum regions around low-dimensional materials. While adding vacuum is effective in reducing short-range interactions between periodic images, it significantly increases the computational cost, especially for large supercells. This is of special importance for MBPT treatment of the excited electronic states with long-range Coulomb interactions. To make these calculations more efficient, efforts have been directed toward implementing Coulomb cutoff methods. Coulomb cutoffs are designed to selectively limit the range of the Coulomb interaction in certain directions, effectively reducing the artificial electrostatic interactions between periodic images without the need for extensive vacuum regions. By truncating the Coulomb potential, these methods ensure that the interaction decays rapidly beyond a certain distance, which allows for the use of smaller unit cells and less vacuum. This reduction not only decreases computational resources but also improves the accuracy of simulations by focusing on the intrinsic properties of the material itself rather than the vacuum-induced artifacts. \\
\noindent In particular, two Coulomb cutoff techniques were implemented in Quantum ESPRESSO for both 1D \cite{coulomb1D} and 2D \cite{coulomb2D} systems. These methods have proven to be highly effective in accelerating the convergence of ground-state properties --- such as the total energy, the Fermi level, and the phonon dispersion --- without the need for excessively large vacuum regions. However they are less effective in the calculation of the energy dispersion, particularly for empty states. This limitation is especially pronounced in 1D systems, where achieving convergence of the band structure with respect to the vacuum region is virtually impossible. \\
\noindent Here, we present the prototypical study of MgN$_2$, which represents a particularly pathological case. The relaxed structure and the converged DFT electronic band structure are displayed in Fig.~\ref{mgn2_1}.
For the DFT part, the QE integrated suite \cite{giannozzi2009quantum, giannozzi2017advanced} was used. A norm-conserving, scalar-relativistic pseudopotential from the PseudoDojo repository (v0.4) \cite{PseudoDojo} was employed, using GGA-PBE XC functional \cite{PBE}. Upon convergence, a kinetic energy cutoff of 80 Ry was chosen. A uniform Monkhorst-Pack $\mathbf{k}$-point mesh with dimension of $1\times1\times24$ was employed. To prevent interaction between periodic replicas, a minimum vacuum region of 16 \si{\angstrom} along the non-periodic ($xy$) directions was introduced. Structural relaxation was considered converged when the maximum component of the residual ionic forces dropped below $10^{-8}$ Ry/Bohr. MgN$_2$ unit cell (Fig.~\ref{mgn2_1}a), with a lattice parameter of 3.823 \si{\angstrom}, is composed of a single Mg atom bonded with two N atoms with an angle of about 38°, forming a triangle in the bonding plane. MgN$_2$ is a direct-gap semiconducting wire, with a calculated narrow-gap of 0.26 eV at the $\Gamma$ point (Fig.~\ref{1D_bands}a). Interestingly, the extremely flat conduction band is separated from the next empty band by an energy gap ($\approx 1.2$ eV at the $\Gamma$ point) that is larger than the electronic gap itself. This gap is even greater for the valence band relative to the closest occupied band, measuring approximately 5 eV at the Z point. These substantial energy separations suggest that interactions or mixing between these bands are minimal and that they could be associated with different atomic orbitals or electronic configurations that are well separated in energy. In Fig.~\ref{VACUUM_1} we present the band structure of MgN$_2$, calculated with two different vacuum regions: \SI{16}{\angstrom} (cyan) and \SI{18}{\angstrom} (blue). Even with a small increase of just \SI{2}{\angstrom}, the electronic bands are well converged only up to few eV above the Fermi level (red dashed line).
This difficulty can be attributed to the reduced dimensionality of 1D systems. In these systems, electronic screening is significantly less effective compared to 2D or bulk materials, leading to stronger Coulomb interactions. As a result, the electron wavefunctions become highly delocalized into the vacuum region, making it challenging to isolate the system's intrinsic properties. This issue significantly hinders the calculation of excited-state properties in 1D systems. Indeed, most computational codes available for simulating such properties in periodic materials rely on summations over empty states to compute key quantities like the correlation self-energy $\Sigma_c$. Therefore, the inability to properly converge the band structure of 1D materials due to insufficient vacuum can lead to inaccuracies in predicting excited-state phenomena. To bypass the issue, we employed the following strategy: 
\begin{itemize}
    \item The QP corrections to the band gap were calculated simultaneously for varying amounts of vacuum.
    \item For each vacuum level, the summations over empty states were truncated at an increasing energy threshold.
\end{itemize}
\noindent The reference for the zero energy should be chosen accordingly, preferably starting from the Fermi level. An example of this procedure is shown in Fig.~\ref{VACUUM_2} for MgN$_2$. As the vacuum increases, the number of empty states within each energy threshold also increases, leading to a significantly higher computational cost at each step. Our calculations indicate that a vacuum of 15–16 \si{\angstrom} and an energy threshold of $20$ eV (from the Fermi level) are generally sufficient to converge the correction to the band gap, resulting in a discrepancy of less than $1\%$ compared to a vacuum of up to \SI{30}{\angstrom}. This ``recipe'' allows for simultaneous control of the convergence with respect to both vacuum and empty states in MBPT calculations, with the QP correction to the band gap serving as the convergence parameter, bypassing the direct dependence of the results on these quantities while reducing the computational cost. \\

\clearpage
\section{DFT-PBE electronic structures}\label{DFTPBE_band}
\begin{figure*}[h!]
  \centering
  \begin{tabular}{@{}c@{}}
    \includegraphics[scale=0.4]{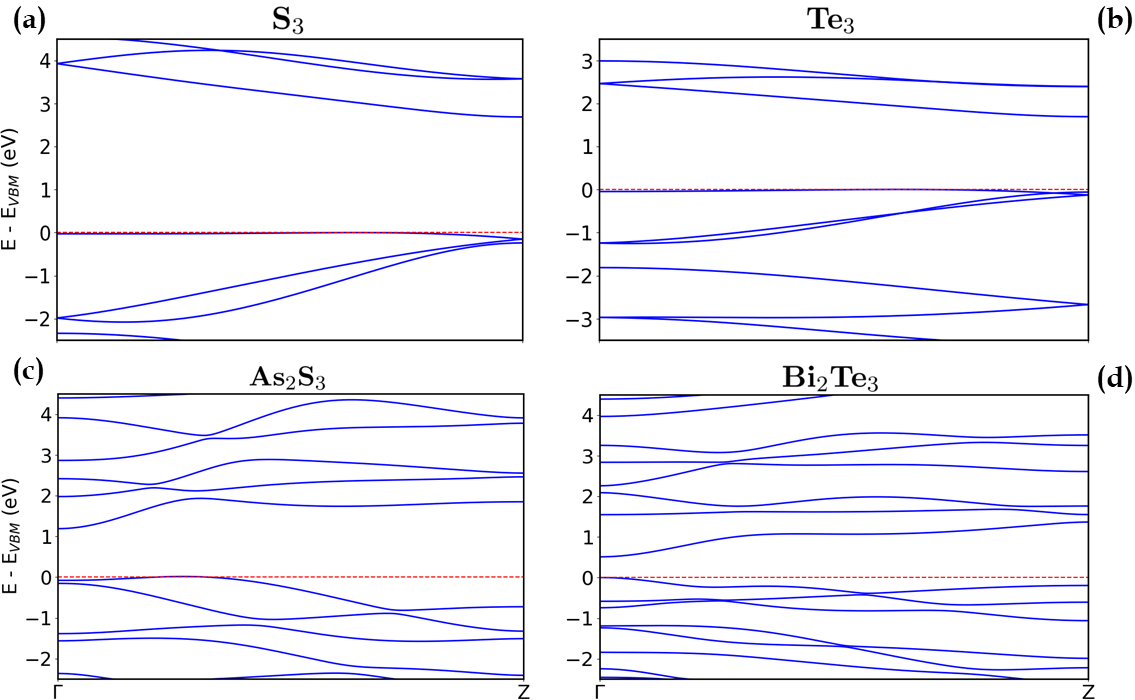} \\[\abovecaptionskip]
  \end{tabular}
  \caption{Electronic band structures of $(\mathbf{a})$ $\mathrm{S}_3$, $(\mathbf{b})$ $\mathrm{Te}_3$, $(\mathbf{c})$ $\mathrm{As}_2\mathrm{S}_3$ and $(\mathbf{d})$ $\mathrm{Bi}_2\mathrm{Te}_3$, calculated at the DFT level, using a GGA-PBE XC functional. Energy zero is set as the top of the valence bands. The zero level is highlighted with a red dotted line. \label{1D_bands}}
\end{figure*}

\begin{figure*}[h!]
  \centering
  \begin{tabular}{@{}c@{}}
    \includegraphics[scale=0.4]{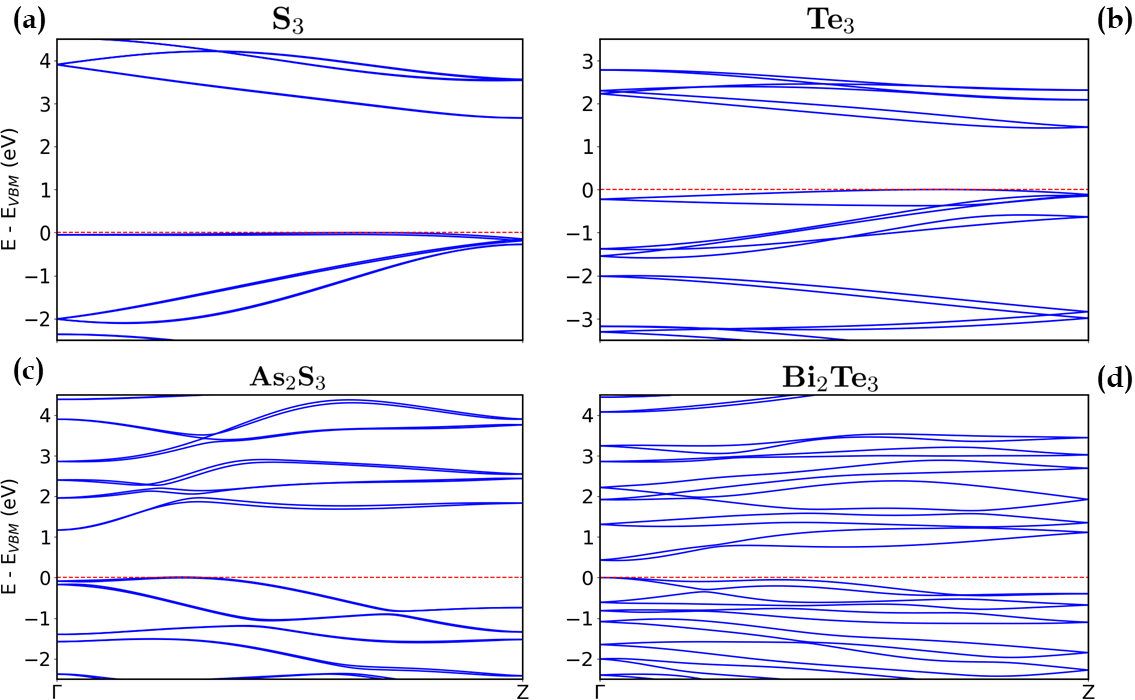} \\[\abovecaptionskip]
  \end{tabular}
  \caption{Electronic band structures of $(\mathbf{a})$ $\mathrm{S}_3$, $(\mathbf{b})$ $\mathrm{Te}_3$, $(\mathbf{c})$ $\mathrm{As}_2\mathrm{S}_3$ and $(\mathbf{d})$ $\mathrm{Bi}_2\mathrm{Te}_3$, calculated at the DFT level, using a GGA-PBE XC functional. SOC was included. Energy zero is set as the top of the valence bands. The zero level is highlighted with a red dotted line. \label{1D_bands_soc}}
\end{figure*}

In this section we present the electronic band structures of the wires at DFT-PBE level without and with the inclusion of SOC, respectively, in Figs.~\ref{1D_bands}-\ref{1D_bands_soc}. We limit the detailed discussion on the SOC case. As pointed out in the main text, all four wires possess an indirect-gap nature, often presenting a small difference between the direct and indirect electronic gap. The corresponding values are listed in Table~1 in the main text. \\
\noindent S$_3$ has an estimated DFT indirect electronic gap of 2.66 eV (Fig.~\ref{1D_bands}a), with a direct gap of 2.76 eV.
The CBM is located at the high-symmetry Z point, while the VBM is located between the $\Gamma$ and Z points of the SOC-split flat valence band. The band structure of Te$_3$ shows many similarities with that of S$_3$, with an estimated indirect electronic gap of 1.43 eV (Fig.~\ref{1D_bands}b). 
Both the VBM and the CBM are located near the high-symmetry Z point. The direct electronic gap is only 40 meV greater and it is located close to the Z point. As$_2$S$_3$ and Bi$_2$Te$_3$ exhibit similar structures and band structures, and both are significantly influenced by SOC. As$_2$S$_3$ possesses an indirect band gap of 1.16 eV (Fig.~\ref{1D_bands}c), while the direct gap was estimated to be 1.26 eV.
In this material, the CBM is located at the $\Gamma$ point, while the VBM is situated between the $\Gamma$ and Z points. In contrast, Bi$_2$Te$_3$ is a nearly-direct narrow-gap semiconductor with an electronic band gap of 0.419 eV (Fig.~\ref{1D_bands}d).
Here, the VBM is located at the $\Gamma$ point, and the CBM is very close to the $\Gamma$ point. Indeed, the direct band gap at $\Gamma$ is only 2 meV larger than the indirect band gap. \\

\clearpage
\section{MBPT supporting data}\label{SIFigures}

\begin{table*}[htbp!]
\small
\renewcommand{\arraystretch}{1.5}
\caption{Calculated lowest direct electronic band gaps $(E_g)$, at the G$_0$W$_0$ and $\mathrm{ev}GW$ level, together with the BSE optical gaps $(E^{\mathrm{BSE}}_{opt})$ and the corresponding binding energies $(E^{\mathrm{BSE}}_b)$ and radii $(r^{\mathrm{BSE}}_b)$ of the lowest bright excitons. In the case of an indirect band gap, the corresponding direct band gap is also reported in square brackets. The reported values are expressed in eV, except for the radius which is expressed in \si{\angstrom}. SOC and semi-core corrections were included.}
\setlength{\tabcolsep}{1pt}
\begin{tabularx}{\textwidth}{>{\centering\arraybackslash}p{1.1cm}>{\centering\arraybackslash}p{1.4cm}>{\centering\arraybackslash}p{2.0cm}>{\centering\arraybackslash}p{2.0cm}>{\centering\arraybackslash}p{2.0cm}>{\centering\arraybackslash}p{1.1cm}>{\centering\arraybackslash}p{2.3cm}>{\centering\arraybackslash}p{2.1cm}>{\centering\arraybackslash}p{1.8cm}}
    \hline

   & $E^{G_0W_0}_{g}$ & $E^{\mathrm{BSE}/G_0W_0}_{opt}$ & $E^{\mathrm{BSE}/G_0W_0}_{b}$ & $r^{\mathrm{BSE}/G_0W_0}_b$ & $E^{\mathrm{ev}GW}_{g}$ & $E^{\mathrm{BSE}/\mathrm{ev}GW}_{opt}$ & $E^{\mathrm{BSE}/\mathrm{ev}GW}_{b}$ & $r^{\mathrm{BSE}/\mathrm{ev}GW}_b$ \\ \hline
    \textbf{S}$_3$ & 5.46 & 3.62 & 1.93 & 12.5 & 6.25 & 4.07 & 2.27 & 12.4 \\
    & [5.55] & & & & [6.34] & & & \\
    \textbf{Te}$_3$ & 3.47 & 1.85 & 1.65 & 14.8 & 4.27 & 2.17 & 2.14 & 14.7 \\
    & [3.50] & & & & [4.31] & & & \\
    \textbf{As}$_2$\textbf{S}$_3$ & 2.26 & 1.41 & 0.89 & 12.00 & 3.00 & 2.71 & 0.33 & 14.4 \\
    & [2.30] & & & & [3.04] & & & \\
    \textbf{Bi}$_2$\textbf{Te}$_3$ & 1.54 & 0.77 & 0.77 & 13.2 & 1.65 & 0.80 & 0.85 & 13.2 \\
    \hline
\end{tabularx}
\label{tab3_1D}
\end{table*}
\clearpage
\subsection{Absorption spectra and electronic band structures at G$_0$W$_0$ and G$_0$W$_0$/BSE level}\label{app:g0w0}

\begin{figure*}[h!]
  \centering
  \begin{tabular}{cc}
    \includegraphics[scale=0.45]{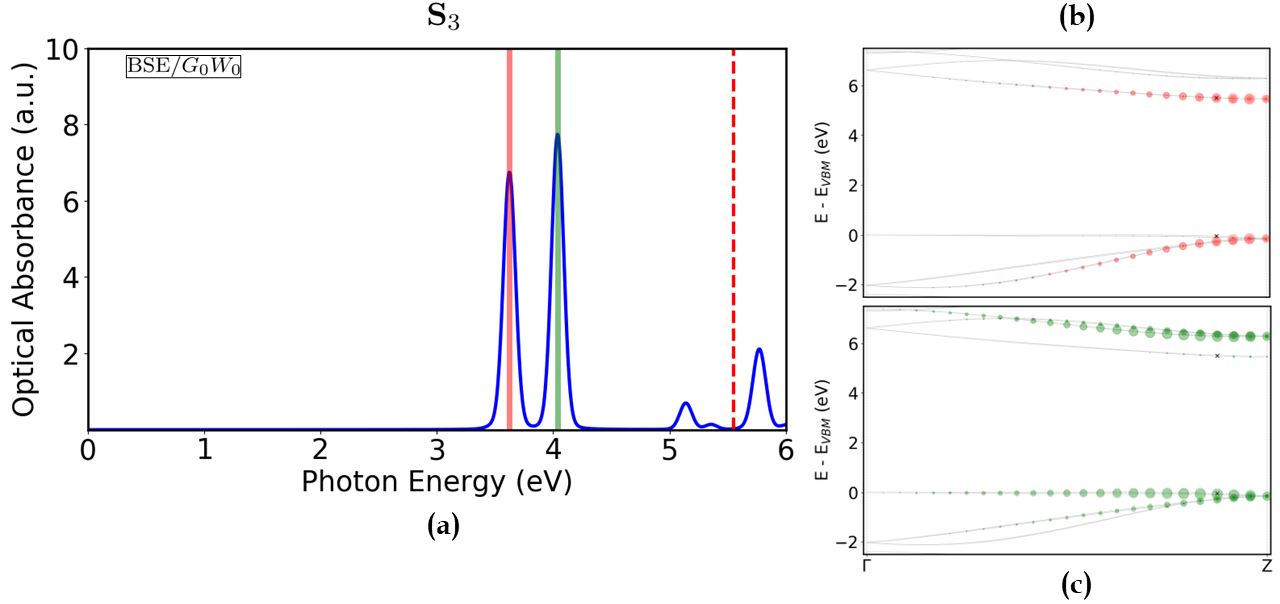} \\
    \includegraphics[scale=0.45]{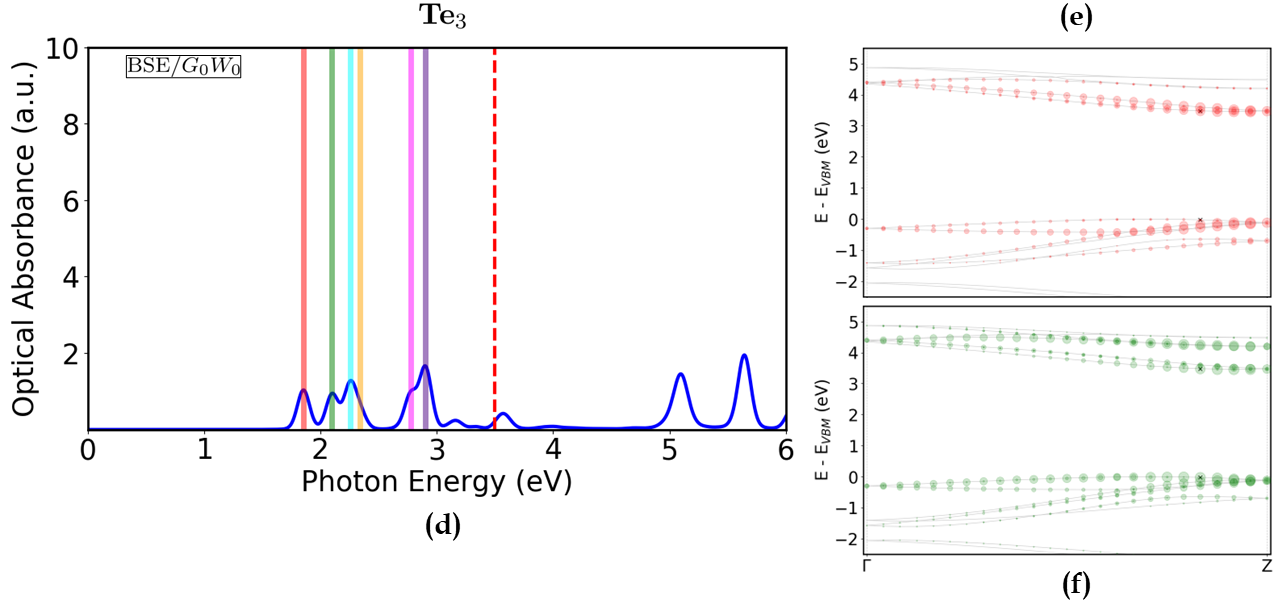}
  \end{tabular}
  \caption{[Left] Absorption spectra (solid blue) of $(\mathbf{a})$ $\mathrm{S}_3$ and $(\mathbf{d})$ $\mathrm{Te}_3$, expressed in terms of the optical absorbance $A(\omega)$, calculated at the $G_0W_0/$BSE level. The corresponding $G_0W_0-$corrected direct electronic band gaps (dashed red) are shown as a reference. A broadening of 50 meV was used. [Right] Electronic band structures (solid grey), calculated at the $G_0W_0$ level, of $\mathrm{S}_3$ $(\mathbf{b}-\mathbf{c})$ and $\mathrm{Te}_3$ $(\mathbf{e}-\mathbf{f})$. The colored dots (red and green) represent the single-particle transitions contributing to the first two bright excitons, and their size is proportional to the intensity of the transition - renormalized to the highest value. The corresponding excitonic peaks are highlighted in the relative absorption spectra (solid red and green), together with other meaningful higher excitations below the electronic gap (see Fig.~\ref{Te3_exc_bands_g0w0} of Appendix B). Energy zero is set as the top of the valence bands. SOC and semi-core corrections were included.\label{1D_G0W0BSE_1}}
\end{figure*}
\begin{figure*}[h!]
  \centering
  \begin{tabular}{cc}
    \includegraphics[scale=0.45]{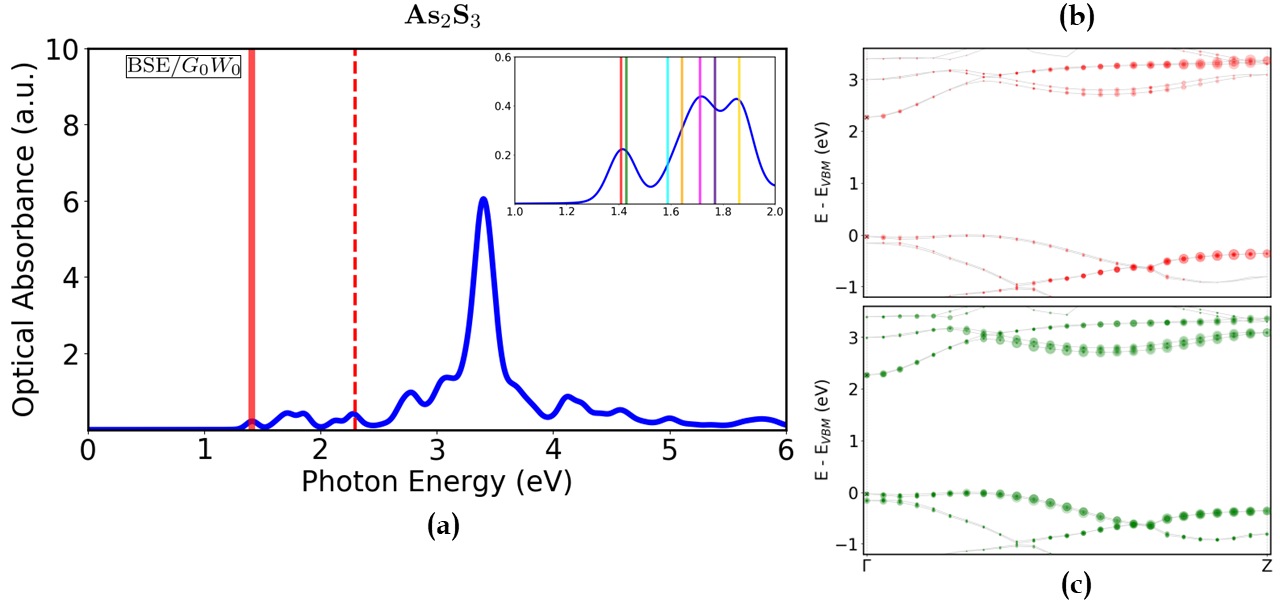} \\
    \includegraphics[scale=0.45]{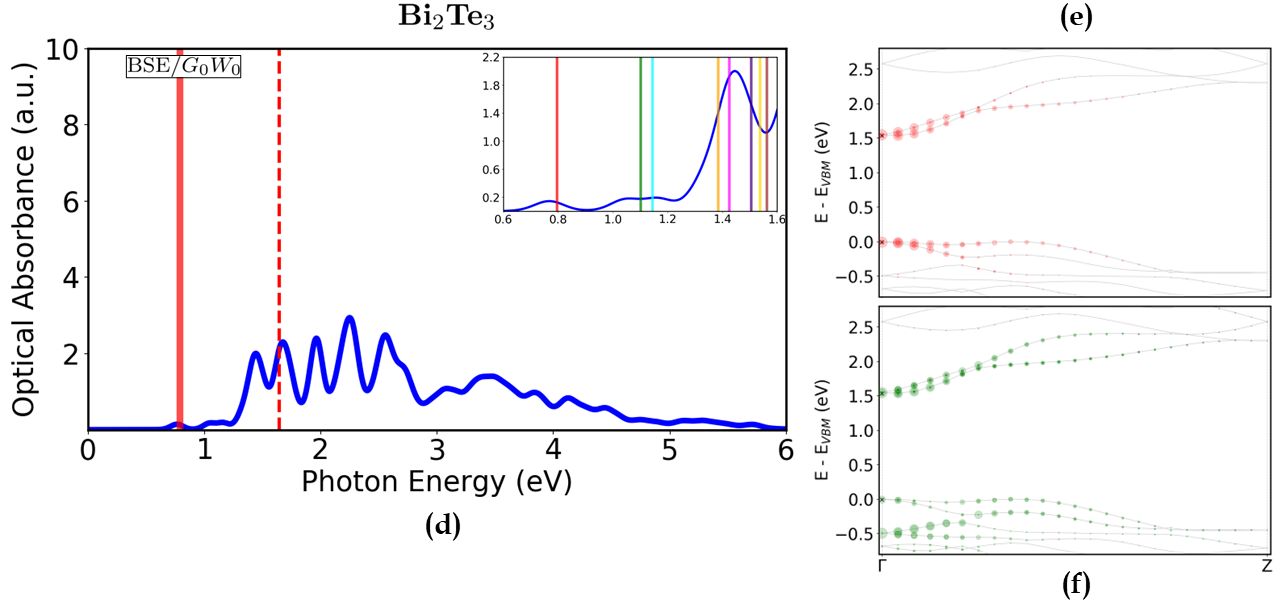}
  \end{tabular}
  \caption{[Left] Absorption spectra (solid blue) of $(\mathbf{a})$ $\mathrm{As}_2\mathrm{S}_3$ and $(\mathbf{d})$ $\mathrm{Bi}_2\mathrm{Te}_3$, expressed in terms of the optical absorbance $A(\omega)$, calculated at the $G_0W_0/$BSE level. The corresponding $G_0W_0-$corrected direct electronic band gaps (dashed red) are shown as a reference. A broadening of 50 meV was used. [Right] Electronic band structures (solid grey), calculated at the $G_0W_0$ level, of $\mathrm{As}_2\mathrm{S}_3$ $(\mathbf{b}-\mathbf{c})$ and $\mathrm{Bi}_2\mathrm{Te}_3$ $(\mathbf{e}-\mathbf{f})$. The colored dots (red and green) represent the single-particle transitions contributing to the first two bright excitons, and their size is proportional to the intensity of the transition - renormalized to the highest value. The corresponding excitonic peaks are highlighted in the relative absorption spectra (solid red and green) and in their insets, together with other meaningful higher excitations below the electronic gap (see Figs.~\ref{As2S3_exc_bands_g0w0}-\ref{Bi2Te3_exc_bands_g0w0}). Energy zero is set as the top of the valence bands. SOC and semi-core corrections were included.\label{1D_G0W0BSE_2}}
\end{figure*}
\begin{figure*}[h!]
  \centering
  \begin{tabular}{@{}c@{}}
    \includegraphics[scale=0.5]{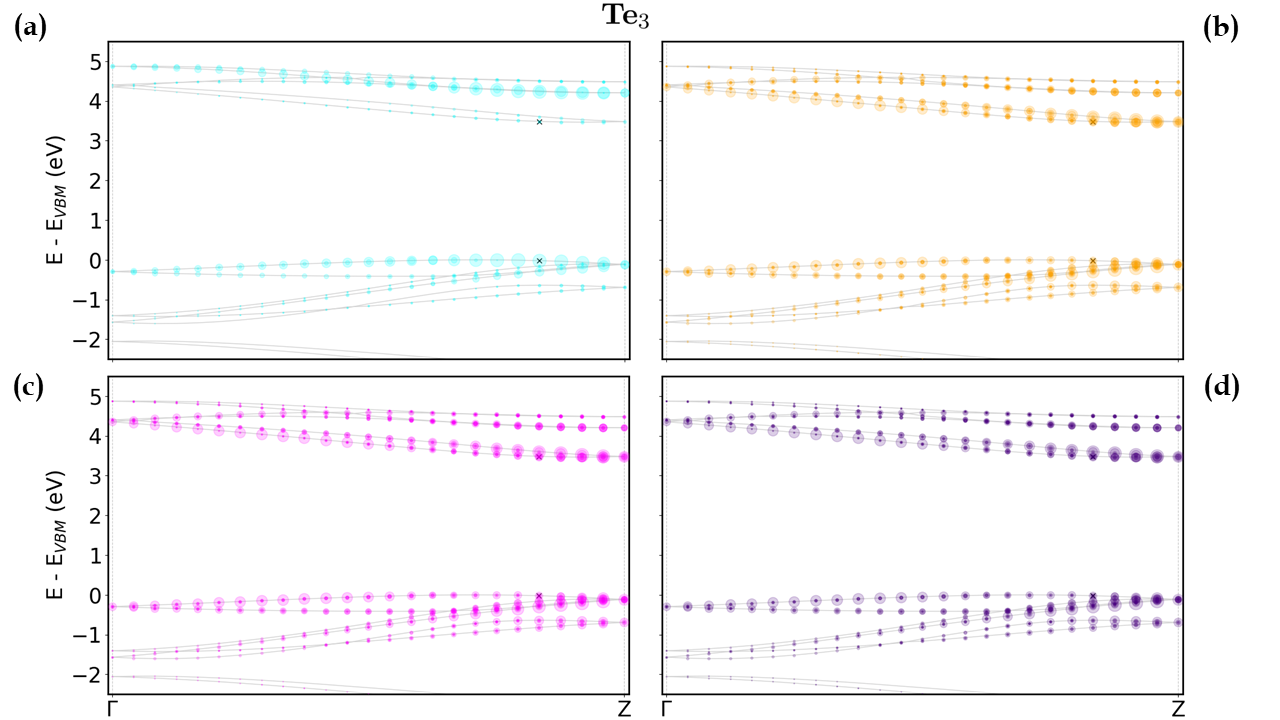} \\[\abovecaptionskip]
  \end{tabular}
  \caption{Electronic band structures (solid grey), calculated at the $G_0W_0$ level, of $\mathrm{Te}_3$. The colored dots (cyan $(\mathbf{a})$, orange $(\mathbf{b})$, magenta $(\mathbf{c})$ and indigo $(\mathbf{d})$) represent the single-particle transitions contributing to the corresponding bright excitons highlighted in the relative absorption spectrum (Fig.~\ref{1D_G0W0BSE_1} of Appendix B), and their size is proportional to the intensity of the transition - renormalized to the highest value. Energy zero is set as the top of the valence bands. SOC and semi-core corrections were included.\label{Te3_exc_bands_g0w0}}
\end{figure*}
\begin{figure*}[h!]
  \centering
  \begin{tabular}{@{}c@{}}
    \includegraphics[scale=0.48]{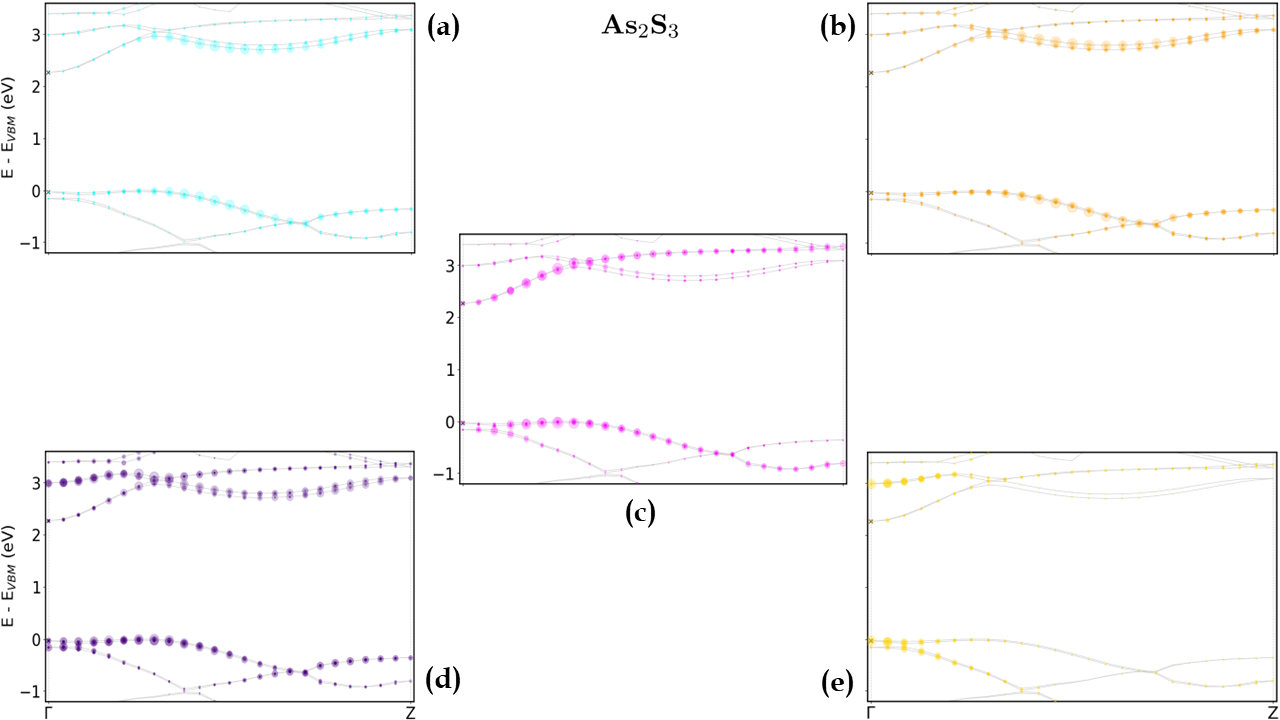} \\[\abovecaptionskip]
  \end{tabular}
  \caption{Electronic band structures (solid grey), calculated at the $G_0W_0$ level, of $\mathrm{As}_2\mathrm{S}_3$. The colored dots (cyan $(\mathbf{a})$, orange $(\mathbf{b})$, magenta $(\mathbf{c})$, indigo $(\mathbf{d})$ and gold $(\mathbf{e})$) represent the single-particle transitions contributing to the corresponding bright excitons highlighted in the relative absorption spectrum (Fig.~\ref{1D_G0W0BSE_2}), and their size is proportional to the intensity of the transition - renormalized to the highest value. Energy zero is set as the top of the valence bands. SOC and semi-core corrections were included.\label{As2S3_exc_bands_g0w0}}
\end{figure*}

\begin{figure*}[h!]
  \centering
  \begin{tabular}{@{}c@{}}
    \includegraphics[scale=0.48]{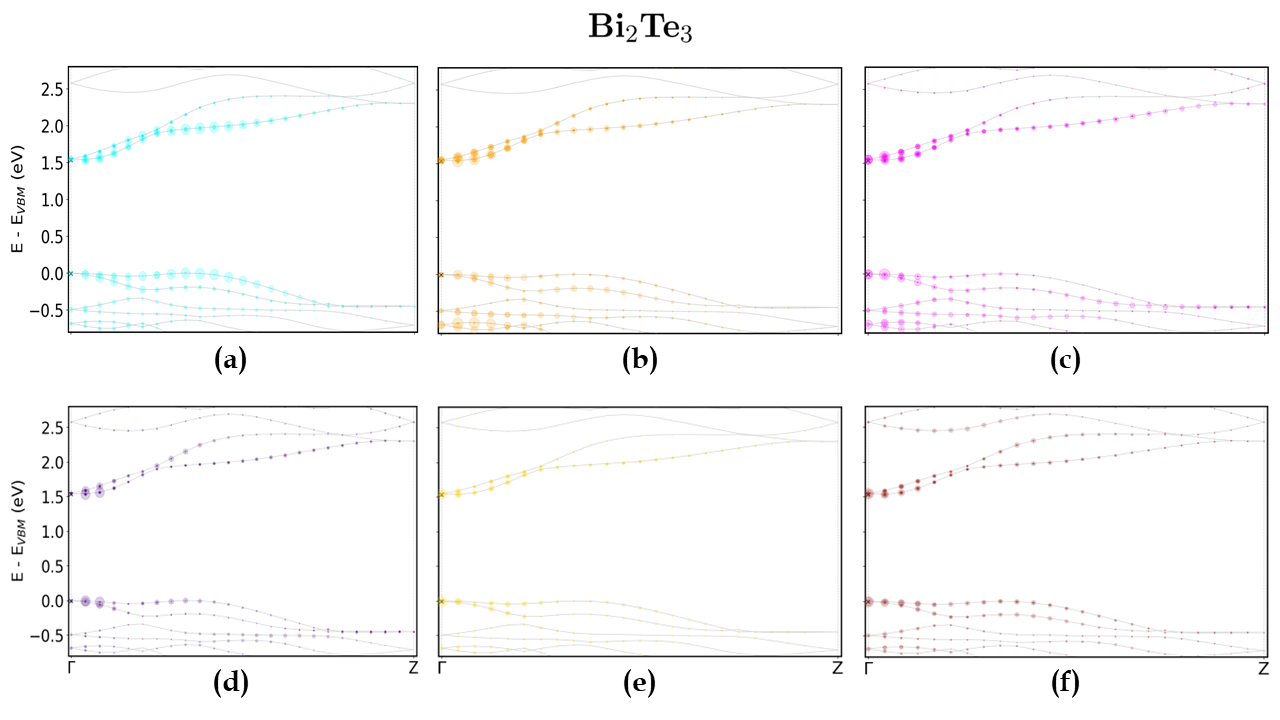} \\[\abovecaptionskip]
  \end{tabular}
  \caption{Electronic band structures (solid grey), calculated at the $G_0W_0$ level, of $\mathrm{Bi}_2\mathrm{Te}_3$. The colored dots (cyan $(\mathbf{a})$, orange $(\mathbf{b})$, magenta $(\mathbf{c})$, indigo $(\mathbf{d})$, gold $(\mathbf{e})$ and brown $(\mathbf{f})$) represent the single-particle transitions contributing to the corresponding bright excitons highlighted in the relative absorption spectrum (Fig.~\ref{1D_G0W0BSE_2}), and their size is proportional to the intensity of the transition - renormalized to the highest value. Energy zero is set as the top of the valence bands. SOC and semi-core corrections were included.\label{Bi2Te3_exc_bands_g0w0}}
\end{figure*}
\begin{figure*}[h!]
  \centering
  \begin{tabular}{@{}c@{}}
    \includegraphics[scale=0.4]{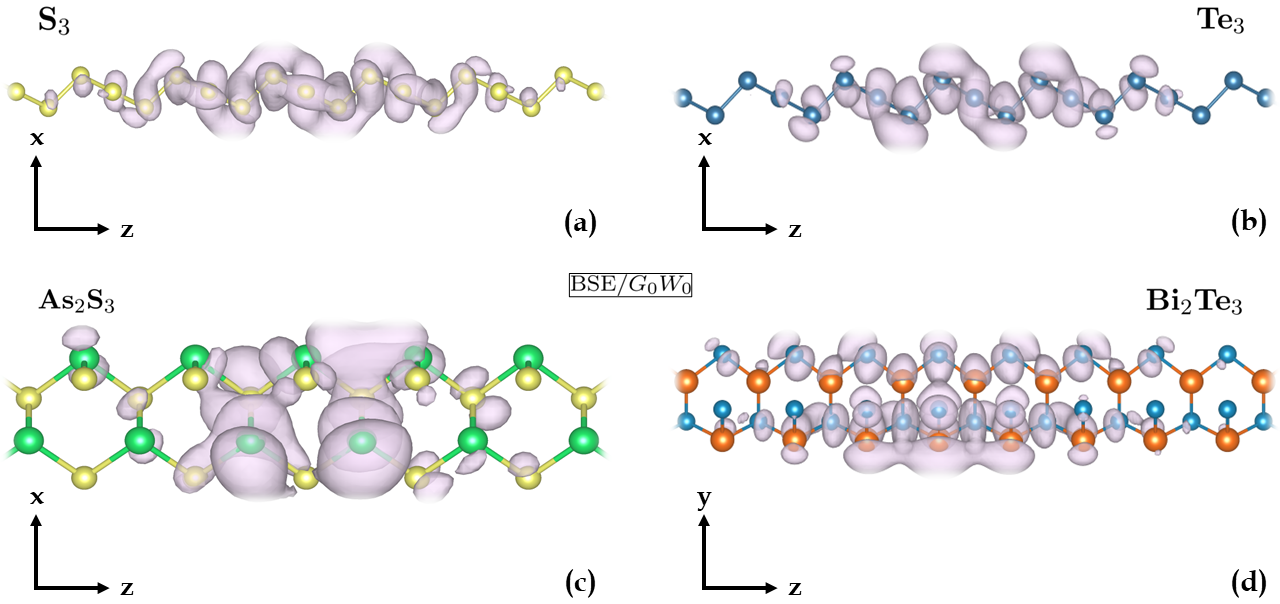} \\[\abovecaptionskip]
  \end{tabular}
  \caption{Plots in direct space of the excitonic wavefunctions of the first bright exciton of $(\mathbf{a})$ $\mathrm{S}_3$, $(\mathbf{b})$ $\mathrm{Te}_3$, $(\mathbf{c})$ $\mathrm{As}_2\mathrm{S}_3$ and $(\mathbf{d})$ $\mathrm{Bi}_2\mathrm{Te}_3$, calculated at the $G_0W_0/$BSE level. The position of the hole was chosen based on the localization of the valence electrons contributing to the exciton. The number of cell repetitions in the periodic direction (7, 7, 5, 8) was increased until the wavefunction decayed to zero.\label{1D_EXC_g0w0}}
\end{figure*}
\clearpage
\subsection{Electronic band structures at evGW level} \label{app:evgw}

\begin{figure*}[h!]
  \centering
  \begin{tabular}{@{}c@{}}
    \includegraphics[scale=0.48]{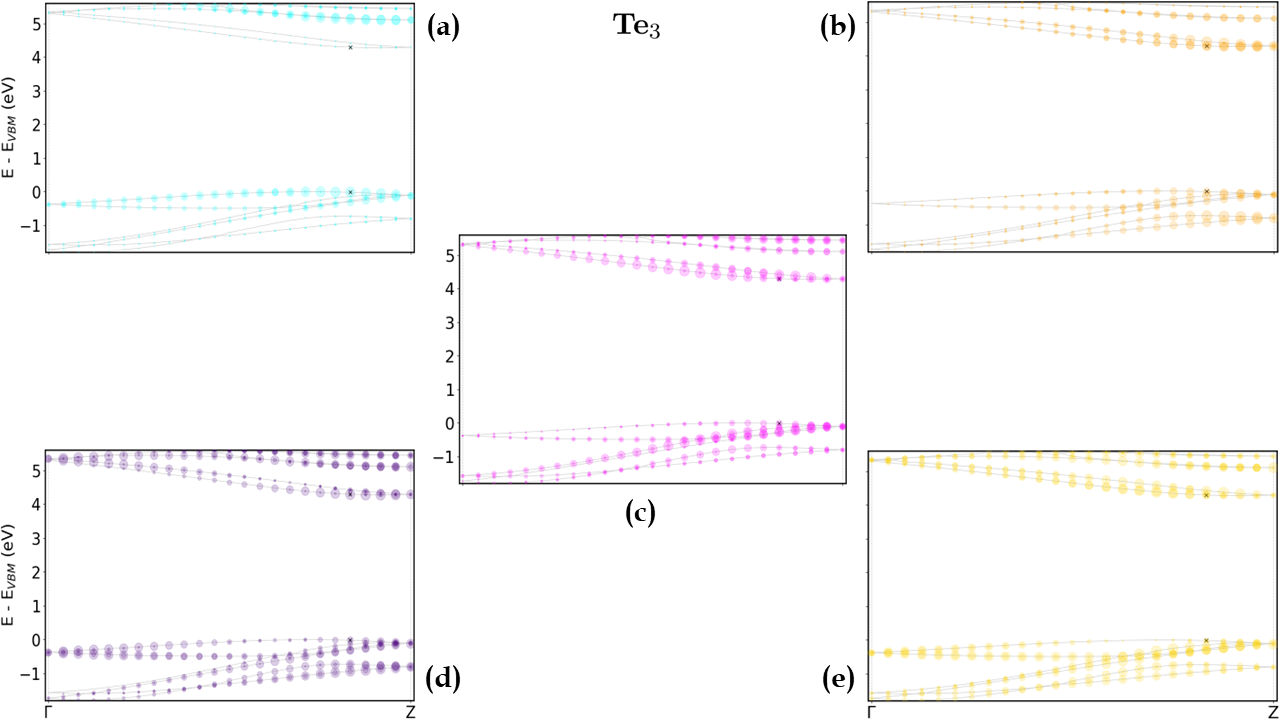} \\[\abovecaptionskip]
  \end{tabular}
  \caption{Electronic band structures (solid grey), calculated at the ev$GW$ level, of $\mathrm{Te}_3$. The colored dots (cyan $(\mathbf{a})$, orange $(\mathbf{b})$, magenta $(\mathbf{c})$, indigo $(\mathbf{d})$ and gold $(\mathbf{e})$) represent the single-particle transitions contributing to the corresponding bright excitons highlighted in the relative absorption spectrum (Fig.~\ref{1D_G0W0BSE_1}), and their size is proportional to the intensity of the transition - renormalized to the highest value. Energy zero is set as the top of the valence bands. SOC and semi-core corrections were included.\label{Te3_exc_bands_evgw}}
\end{figure*}

\begin{figure*}[h!]
  \centering
  \begin{tabular}{@{}c@{}}
    \includegraphics[scale=0.48]{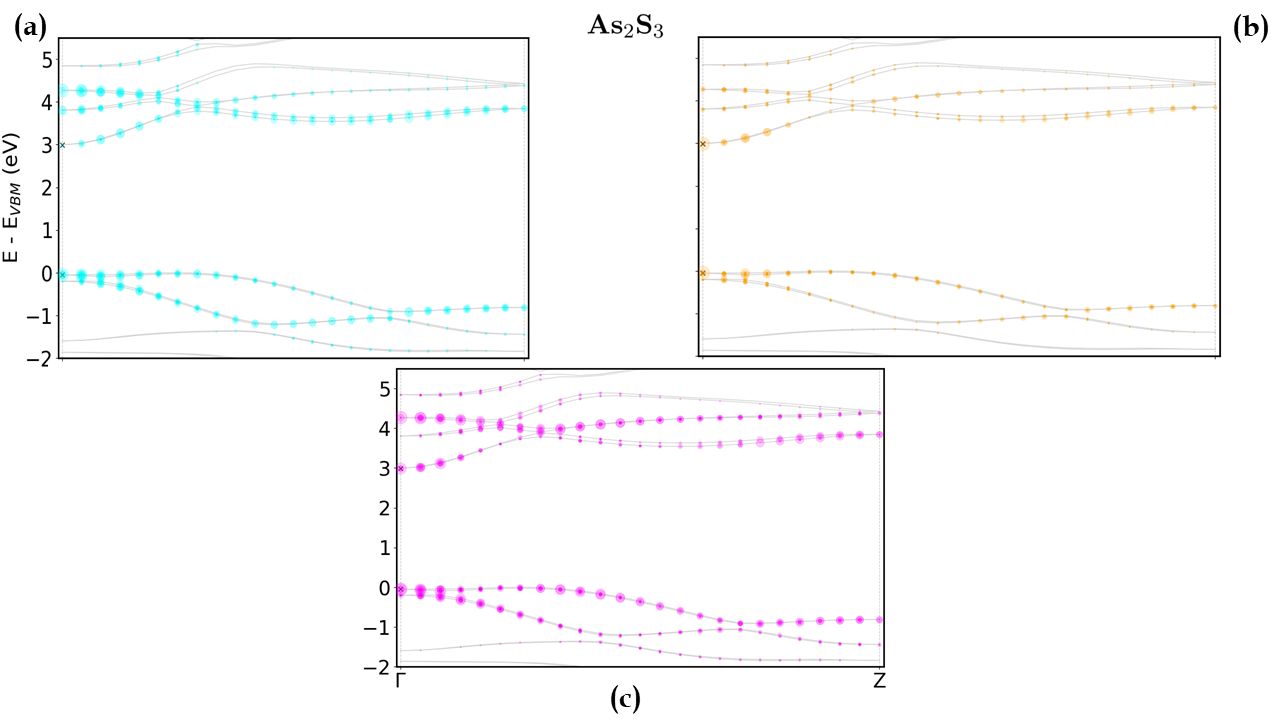} \\[\abovecaptionskip]
  \end{tabular}
  \caption{Electronic band structures (solid grey), calculated at the ev$GW$ level, of $\mathrm{As}_2\mathrm{S}_3$. The colored dots (cyan $(\mathbf{a})$, orange $(\mathbf{b})$ and magenta $(\mathbf{c})$) represent the single-particle transitions contributing to the corresponding bright excitons highlighted in the relative absorption spectrum (Fig.~7 of the main text), and their size is proportional to the intensity of the transition - renormalized to the highest value. Energy zero is set as the top of the valence bands. SOC and semi-core corrections were included.\label{As2S3_exc_bands_evgw}}
\end{figure*}

\begin{figure*}[h!]
  \centering
  \begin{tabular}{@{}c@{}}
    \includegraphics[scale=0.48]{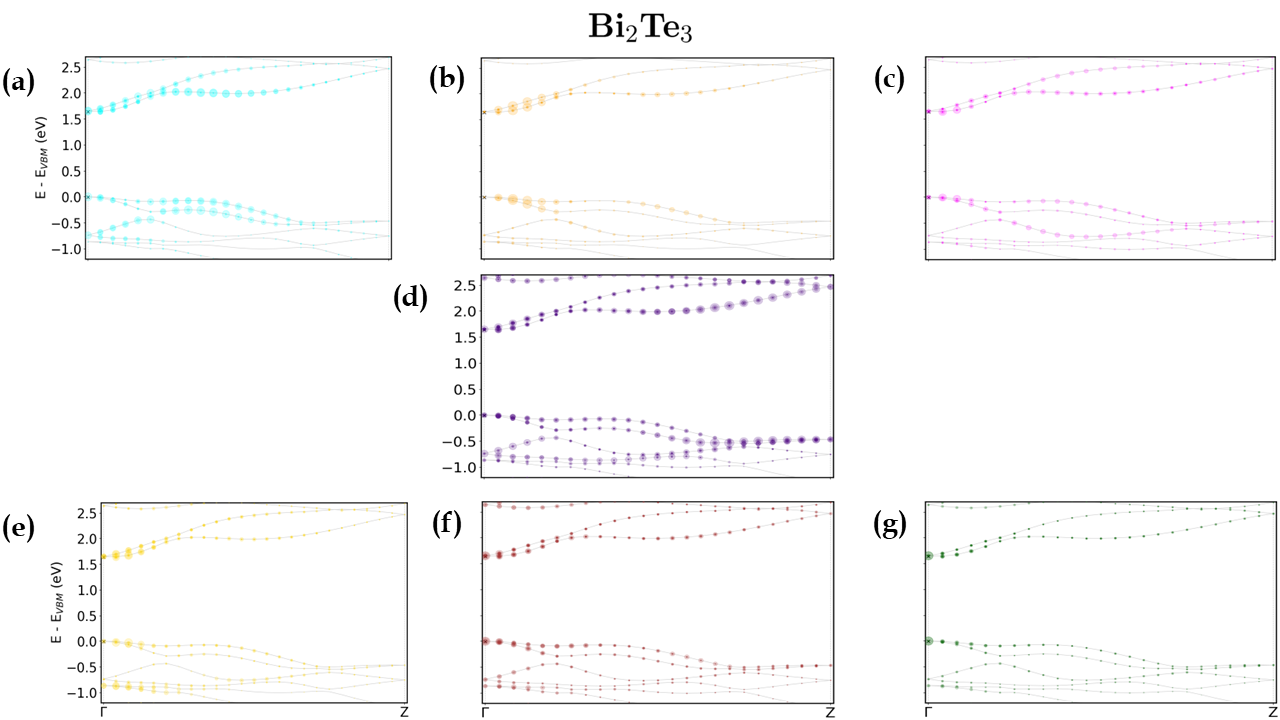} \\[\abovecaptionskip]
  \end{tabular}
  \caption{Electronic band structures (solid grey), calculated at the ev$GW$ level, of $\mathrm{Bi}_2\mathrm{Te}_3$. The colored dots (cyan $(\mathbf{a})$, orange $(\mathbf{b})$, magenta $(\mathbf{c})$, indigo $(\mathbf{d})$, gold $(\mathbf{e})$, brown $(\mathbf{f})$ and dark green $(\mathbf{g})$) represent the single-particle transitions contributing to the corresponding bright excitons highlighted in the relative absorption spectrum (Fig.~7 of the main text), and their size is proportional to the intensity of the transition - renormalized to the highest value. Energy zero is set as the top of the valence bands. SOC and semi-core corrections were included.\label{Bi2Te3_exc_bands_evgw}}
\end{figure*}
\clearpage
\section*{References}
\providecommand{\latin}[1]{#1}
\makeatletter
\providecommand{\doi}
  {\begingroup\let\do\@makeother\dospecials
  \catcode`\{=1 \catcode`\}=2 \doi@aux}
\providecommand{\doi@aux}[1]{\endgroup\texttt{#1}}
\makeatother
\providecommand*\mcitethebibliography{\thebibliography}
\csname @ifundefined\endcsname{endmcitethebibliography}
  {\let\endmcitethebibliography\endthebibliography}{}